\documentclass[twocolumn]{aastex701}

\usepackage{graphicx}
\usepackage{txfonts}
\usepackage{multirow}
\usepackage{booktabs}
\usepackage[dvipsnames]{xcolor}     
\usepackage{booktabs}       
\usepackage{caption}       
\usepackage{array}         
\usepackage{float}

\newcommand{\teff}{$T_{\rm eff}$ }

\definecolor{meridithgreen}{RGB}{0, 150, 0}

\begin{document}

\title{Stranger Things: A Grid-based Survey of Strange Modes in Post-Main Sequence Models}

\author[orcid=0000-0003-3759-7616,gname='D\'ora',sname='\'ora',sname='Tarczay-Neh\'z']{D. Tarczay-Nehéz}
\affiliation{Konkoly Observatory, HUN-REN CSFK, MTA Centre of Excellence, Budapest, Konkoly Thege Mikl\'os \'ut 15-17, Hungary}
\affiliation{MTA–HUN-REN CSFK Lendület "Momentum" Stellar Pulsation Research Group}
\email[show]{tarczaynehez.dora@csfk.org}  

\author[0000-0002-8159-1599]{L\'{a}szl\'{o} Moln\'{a}r}
\affiliation{Konkoly Observatory, HUN-REN CSFK, MTA Centre of Excellence, Budapest, Konkoly Thege Mikl\'os \'ut 15-17, Hungary}
\affiliation{MTA–HUN-REN CSFK Lendület "Momentum" Stellar Pulsation Research Group}
\affiliation{ELTE E\"otv\"os Lor\'and University, Institute of Physics and Astronomy, 1117, P\'azm\'any P\'eter s\'et\'any 1/A, Budapest, Hungary}
\email[]{}  

\author[0000-0002-8717-127X]{Meridith Joyce}
\affiliation{Konkoly Observatory, HUN-REN CSFK, MTA Centre of Excellence, Budapest, Konkoly Thege Mikl\'os \'ut 15-17, Hungary}
\affiliation{MTA–HUN-REN CSFK Lendület "Momentum" Stellar Pulsation Research Group}
\affiliation{University of Wyoming, 1000 E University Ave, Laramie, WY USA}
\email[]{}  

\begin{abstract}

We present a systematic survey of strange mode pulsations in Cepheids using \texttt{MESA} and for the linear stability analysis, \texttt{MESA RSP}. 
Our model grid spans $2$--$15\,M_\odot$ in mass and [Fe/H] $= -0.95$--$0.17$ ($Z = 0.0015$--$0.0200$) in metallicity, with four convective overshoot prescriptions.
Strange modes were identified in a relatively small fraction ($\sim 5$--$12.5\%$) of models, occurring at $n_\mathrm{pg} = 5$--$9$, with $n_\mathrm{pg} = 6$--$7$ as the most frequent radial modes.
No unstable solutions were identified beyond $n_\mathrm{pg} = 9$, in contrast to earlier studies reporting strange modes at $n_\mathrm{pg} = 10 - 12$. 
We quantified the duration of the instability crossing phase ($\tau_\mathrm{IS}$), the strange mode phase ($\tau_\mathrm{s}$), and their ratio $\mathcal{P}_\mathrm{s} = \tau_\mathrm{s} / \tau_\mathrm{IS}$. 
Toward higher masses, both $\tau_\mathrm{IS}$ and $\tau_\mathrm{s}$ decrease, yet their ratio shows no systematic trend with mass in models that include convective core overshoot.
The absolute timescales for strange modes remain short, typically $\tau_\mathrm{s} \sim 10^{4.5}$--$10^{6}$ years, while $\tau_\mathrm{IS}$ is often an order of magnitude shorter, implying that these stars may spend a larger fraction of their life in the strange mode phase than in the instability strip itself.
The extended duration of the strange mode phase may enhance the detectability of strange mode pulsators, 
provided that observational precision is sufficient to capture their low-amplitude variability.
The predicted periods (0.6–6.3 days) are well covered by a single 27‑day TESS sector, making strange mode pulsators potentially detectable with current space-based photometry, although blending with nearby sources may pose challenges.

\end{abstract}



\section{Introduction} 

The term "strange modes" was first introduced by \cite{Coxetal1980} and refers to a class of pulsation modes found in post-AGB stars.
These pulsations were initially thought of as a result of strong non-adiabatic effects, and have a characteristic ultra-low amplitude (ULA) light variation on a millimagnitude scale \citep[see, e.g.,][]{GautschySaio1996, BuchlerandKollath2001, Saio2009}. The possibility of strange modes in hot stars was recently revisited using more modern codes by \citet{Gautschy2025}, too.
However, strange modes are not limited to hot stars. They were identified in AGB star models by \citet{Wood-2014}. \citet{Buchleretal1997} and \citet{BuchlerandKollath2001} showed that strange modes can exist in weakly non-adiabatic stars, like classical Cepheids, or RR~Lyrae type stars, as well, even with the inclusion of turbulent convection into the hydrodynamic models. 
Moreover, they not only exist but may thrive in classical Cepheids.
   
As a Cepheid progenitor evolves from the main sequence (MS) towards the red giant branch (RGB), it undergoes the first crossing of the instability strip (IS), where coherent radial pulsations occur. 
During this phase, the outer regions of the star inflate, resulting in the 
inward shift of the partial ionization zones of H and He. 
In this phase, the star evolves rapidly through the Hertzsprung-Russell diagram (HRD) towards the red giant branch (RGB), spending only a short period of time inside the IS. 
As such, most Cepheids we know are not in the first crossing phase.
In moderate- to high-mass red giant stars ($> 2-3M_{\odot}$), the He core ignites before becoming degenerate. These stars do not experience a He-core flash, and do not descend to the red clump or the horizontal branch: instead, as the core starts to burn He, the star starts to shrink at roughly constant luminosity, while the temperature of the star starts to increase.
This shift towards the bluer colors (hotter temperatures), and back after core He exhaustion, is often referred to as the `blue loop' in the literature \citep{Lauterborn1971, Frickeand1972}. 
The duration of the blue loop (i.e., how far it extends towards the blue parts) and the crossing time depends on different stellar physical parameters---e.g., mass, metallicity, rotation, as well as the nuclear reaction rates and opacities---and the inner structure of the star that is dominated by the internal  processes, such as convection and overshoot (\citealt[see, e.g.,][]{Bonoetal2000,XuandLi2004a,XuandLi2004b,Jinetal2015, Walmswelletal2015, Tangetal2016, Zhaoetal2023}, or the most recent work of \citealt{Andersonetal2025} and \citealt{Stucketal2025}).

If the star traverses far enough in temperature, it crosses the IS again (second and third crossing), where stable pulsations can develop. These are usually limited to low-order radial modes, but could also include higher-order overtone modes, given the right conditions. 
In Cepheids and RR~Lyrae stars, the partial ionization layer of the star acts as a sharp potential barrier, dividing the star into two regions: the inner part and the outer surface layers. 
Thus, pulsation modes can be trapped both in the inner parts and in the outer layer of the star. 
Strange modes develop, where the higher overtone modes are trapped in the outer surface layer (decoupled from the inner parts of the star), causing the pulsations to be dominated by surface oscillations \citep{BuchlerandKollath2001}. 
Models showed that in a typical $5M_\odot$ classical Cepheid star, the strange modes have a node number of $n=10-12$ on the blue edge of the instability strip (and somewhat smaller in terms of node number on the red edge). They also showed that strange modes may form their own instability regions in models hotter than the blue edge of the classical instability strip.

 In the following sections, we present a theoretical parameter study to identify under which physical conditions strange modes may occur, and map their position on the Hertzsprung-Russell and \textit{Gaia} color-magnitude diagrams.
    Section 2 gives a brief overview of the history of strange mode observations.
    Section 3 describes the construction of the full model grid using the open-source 'Modules for Experiments in Stellar Astrophysics' (\texttt{MESA}) software package.
    The Section presents the range of masses, metallicity, and overshoot parameters explored. 
    Section 4 describes the setup of the asteroseismic simulations used to probe pulsational stability across the grid, utilizing the \texttt{GYRE} and the Radial Stellar Pulsation (\texttt{RSP}) module of \texttt{MESA}. 
    In Section 5, we present our results, highlighting the regions of instability and their dependence on key stellar parameters. 
    Finally, Section 6 discusses the results, future prospects, including detectability, and concluding remarks.

\section{Previous detection attempts}

    Unstable strange modes are predicted to have small amplitudes, short periods, and weak surface velocities (typically between 0.1 and 1.0 km/s). Their luminosity variations are in the millimagnitude range, making them challenging to observe with traditional ground-based telescopes. Until now, candidates for ULA Cepheids have been reported in the Large Magellanic Cloud (LMC). 
    Early detections by \citet{Buchleretal2005} identified a handful of ULA and strange mode candidates based on Massive Compact Halo Object (MACHO) and Optical Gravitational Lensing Experiment (OGLE-II, and OGLE-III) photometric data.
    Subsequent analyses by \citet{Soszynskietal2008} expanded the sample, listing over 20 possible ULA Cepheids located near the classical Cepheids in both the color–magnitude and period–luminosity diagrams. 
    \cite{Buchleretal2009} have reported an additional $\sim30$ multiperiodic stars, that may be low amplitude Pop II Cepheids or RV Tauri stars.

    The first space-borne attempt to identify such low-amplitude Cepheid candidates was presented by \citet{Szaboetal2009}, based on Cepheid candidates using data from the Convection, Rotation, and planetary Transits (CoRoT) space telescope observations \citep{CoRoT2006}. 
    They reported a sample of 37 potential candidates in the Milky Way.
    However, only six light curves were published, and the full list of targets was not disclosed. 
    In our recent study, we re-examined the six published CoRoT targets using \textit{Gaia} DR3 \citep{BailerJones2021,Gaiadr3} parallaxes and photometry to derive reddening-free absolute magnitudes \citep{seismolab} and positions in the period–luminosity and color–magnitude diagrams. 
    We also analyzed their light curves using both CoRoT and  Transiting Exoplanet Survey Satellite mission \citep[TESS,][]{Ricker-2015} data, calculating Fourier parameters and tracking long-term phase shifts. 
    Based on these results, we concluded that these six candidates are likely rotationally modulated variables—such as spotted stars or $\alpha^2$ Canum Venaticorum types—that fall outside the expected parameter space for Cepheid pulsation \citep{TarczayNehezetal2023}.

    While strange modes in RR~Lyrae stars have not been directly observed yet, their presence was inferred based on their dynamical interaction with the fundamental mode. Models have shown that the 9th overtone can get into resonance with the fundamental mode, leading to the onset of period-doubling in the mode. This altered dynamical state can also destabilize the first overtone, which then starts to pulsate with low amplitude, outside the double-mode regime \citep{Kollath-2011,Molnar-2012AN}. Both of these effects were first observed in the continuous light curves of RR Lyrae stars collected by the \textit{Kepler} space telescope \citep{Szabo-2010,Molnar-2012ApJ}. 
    
    Given that strange modes are associated with low-amplitude light variation (i.e., on the order of millimagnitudes), space-borne missions with increased photometric precision are essential. 
    The ongoing TESS mission offers continuous, high-cadence monitoring across large portions of the sky, with the sensitivity required to detect millimagnitude-level fluctuations. 
    TESS offers a realistic chance to provide observational evidence for the existence of strange mode pulsations in classical Cepheids.

\section{Model grid}

\begin{table*}[ht]
\centering
\caption{Summary of key input parameters used in the \texttt{MESA} simulations.}
\label{tab:mesa_parameters}
\begin{tabular}{llll}
\hline
\textbf{Parameter} & \textbf{Value / Range} & \textbf{Applied in} & \textbf{Notes} \\
\hline
\multicolumn{4}{l}{\textbf{Stellar composition and initial parameters}} \\[0.5ex]
\hline
Initial mass ($M_\star$) & $2.0$–$15.0\,M_\odot$ & All models & $\Delta M = 0.5\,M_\odot$ \\
Initial metallicity ($Z_{\text{init}}$) & $0.0015$–$0.0200$ & All models & $\Delta Z = 0.0005$ \\
Initial helium fraction ($Y_{\text{init}}$) & $0.256$ & All models & Fixed \\
Initial hydrogen fraction ($X_{\text{init}}$) & $1 - Z - Y$ & All models & Computed dynamically \\
\hline
\multicolumn{4}{l}{\textbf{Mixing and convection physics}} \\[0.5ex]
\hline
Mixing length parameter ($\alpha_{\text{MLT}}$) & $1.8$ & All models & \texttt{MLT\_option = 'TDC'} \\
Semi-convection parameter ($\alpha_s$) & $0.1$ & All models &  \\
Convective boundary criterion & Ledoux & All models & \texttt{use\_Ledoux\_criterion = .true.} \\
Overshoot scheme & Exponential & Sets 1–3 & \texttt{overshoot\_scheme = 'exponential'} \\
Overshoot parameters ($f$, $f_0$) & See Table\,\ref{tab:overshoot} & Sets 1–3 & Varies per set \\
\hline
\multicolumn{4}{l}{\textbf{Nuclear and evolutionary setup}} \\[0.5ex]
\hline
Nuclear network & \texttt{pp\_and\_cno\_extras\_o18\_ne22} & All models & Tracks 19 isotopes \\
Stop condition & TP-AGB onset & All models & \texttt{stop\_at\_phase\_TP\_AGB = .true.} \\
\hline
\multicolumn{4}{l}{\textbf{Opacity sources and microphysics}} \\[0.5ex]
\hline
High-$T$ opacity source & OPAL & All models & \texttt{kap\_file\_prefix = 'a09'} \\
Low-$T$ opacity source & Ferguson et al. (2005) & All models & \texttt{kap\_lowT\_prefix = 'lowT\_fa05\_a09p'} \\
Heavy element mixture & Asplund et al. (2009) & All models & Solar-like abundance ('A09') \\
Carbon/Oxygen opacities & Enabled & All models & \texttt{kap\_CO\_prefix = 'a09\_co'} \\
Type2 opacities & Enabled & All models & \texttt{use\_Type2\_opacities = .true.} \\
\hline

\end{tabular}
\end{table*}

    In this study, we use the stable version 23.05.1 of the open-source 'Modules for Experiments in Stellar Astrophysics' (\texttt{MESA}) software package \citep{Paxton2011,Paxton2013,Paxton2015,Paxton2018,Paxton2019,Jermyn2023}. \texttt{MESA} is a versatile, open-source software instrument, offering various options for physical assumptions that can be adjusted, including chemical composition, convective properties, rotation, mass and mass loss rate, binarity, etc, all of which can significantly influence stellar evolution.  
    Our simulations were built to run 
    single-star evolution from the pre-main-sequence (pre-MS) phase up to the onset of the Thermally Pulsing Asymptotic Giant Branch (TP-AGB) phase, as controlled by the parameter \texttt{stop\_at\_phase\_TP\_AGB} set to \texttt{true}.

    This parameter study involved a total of $4104$ individual stellar evolution models. 
    Based on the average Total CPU Time measured for a single model run (approximately $18,000$ seconds\footnote{The single-model average Total CPU Time was calculated using the sum of the Unix \texttt{user} and \texttt{system} times, which reflects the total processor time spent. Typical runs utilized two CPU cores with an average parallelization factor of $\approx 1.6$.}), the total computational cost for the entire grid is estimated to be over $74$ million CPU seconds, equivalent to approximately $20,500$ CPU hours.

    This section details the specific parameters used in our simulations. 
    Table\,\ref{tab:mesa_parameters} summarizes the model parameters used in this paper.
    For reproducibility, the full setup (input scripts and example files) is archived at a GitHub repository of the first author\footnote{\url{https://github.com/tnehezd/Strange-mode-cepheids}, stable release v1.0.1}.
    Demonstrative example input/output files (including GYRE and RSP inlists) are archived at Zenodo DOI 10.5281/zenodo.17229951\footnote{\url{https://zenodo.org/records/17229951}}.

    \subsection{The Stellar Mass and Metallicity Grid}
    
    In order to filter out the blue loop crossing Cepheid candidates through the whole instability strip, we constructed a grid based on metallicity (\texttt{Z\_init}) and the initial mass (\texttt{initial\_mass}) of the star. The grid covers: 
    
\vspace{1em}
    $M_\star \in \left\{ 2.0\,\ldots\,15.0 \right\}\,\,\, \delta M_\star = 0.5M_\odot$
    
    $Z_\mathbf{init} \in \left\{ 0.0015\,\,\ldots\,\,0.0200\right\}\,\,\,\delta Z = 0.0005$,

\vspace{1em}

    \noindent resulting in a $27 \times 38 = 1026$ mass-metallicity ($M-Z$) model set. For all models, the initial helium mass fraction (\texttt{Y\_initial}) was set uniformly to $0.256$. 
    As a result, the effective helium enrichment ratio (${\Delta Y}/{\Delta Z}$) varies across the metallicity range, reaching values as high as ~5 at low metallicity and decreasing to ~0.4 at solar metallicity.
    The corresponding initial hydrogen mass fraction (\texttt{X\_init}) was determined by $X_{init} = 1 - Z_{init} - Y_{init}$. 
    This approach was adapted from the \texttt{mesa\_blue\_loop} of the \texttt{mesa} test case setup.
    We note that the \texttt{mesa\_blue\_loop} test case is not intended for scientific modeling, thus in our models the applied grid and physics parameters were set specifically for this study rather than relying on the default test configuration.

\subsection{Opacities and Heavy Element Abundance Pattern}
    We used \texttt{MESA} built-in opacity tables to describe the efficiency of the energy transport. For high temperatures, we used Opacity Project at Livermore (OPAL) opacity tables of \cite{IglesiasandRogers1993, Iglesiasetal1996} implicitly via \texttt{kap\_file\_prefix = 'a09'}. For low temperatures, we used opacity tables of \cite{Fergusonetal2005} utilizing \texttt{kap\_lowT\_prefix = 'lowT\_fa05\_a09p'}.
    Both regimes were calculated using a solar-like heavy element abundance by \cite{Asplundetal2009} (\texttt{kap\_file\_prefix = 'a09'}), often referred to as 'A09'. The \texttt{Z\_{base}} parameter was dynamically set to be equivalent to \texttt{Z\_{init}}. 
    We used specific Carbon and Oxygen enriched opacity tables (\texttt{kap\_CO\_prefix = 'a09\_co'}) and Type2 opacity tables (\texttt{use\_Type2\_opacities = .true.}), although their impact on the evolution of the stars is expected to be minimal, as the simulations were terminated at the onset of the TP-AGB phase.

\subsection{Convective Core Overshoot}
    
    The impact of the convective core overshoot was investigated by adopting an exponential overshoot scheme (\texttt{overshoot\_scheme = 'exponential'}). 
    It was applied for the two key regions: the top of the helium-burning core (\texttt{overshoot\_zone\_type(1) = 'burn\_H'}), and the bottom of the "non-burning" region (\texttt{overshoot\_zone\_type(2) = 'none'}). 
       
    Following the idea of \cite{Ziolkowska2024}, we investigated the impact of internal turbulent processes on stellar evolution across the instability strip by running each $M$–$Z$ model in four separate configurations. 
    The fiducial run excluded convective core overshoot entirely, while the remaining three runs implemented distinct overshoot prescriptions using exponential decay schemes. 
    The corresponding decay parameters were set are listed in Table\,\ref{tab:overshoot}.
    Considering the three overshoot sets including the reference model, in total, $4\times 1026 = 4104$ individual stellar evolution models were run.

   In all cases, the convection was described by using the Mixing Length Theory (MLT) with a mixing length parameter \citep{JoyceandTayar2023}, $\alpha_\mathrm{MLT} = 1.8$ \citep{XuandLi2004a,XuandLi2004b}, and the time-dependent convection (\texttt{MLT\_option = 'TDC'}). 
   To identify the convective boundaries, the Ledoux criterion was applied (\texttt{use\_Ledoux\_criterion = .true.}), with the semi-convection parameter (\texttt{alpha\_semiconvection }), $\alpha_{s} = 0.1$. 
   This follows the parameters set in \texttt{mesa\_blue\_loop} test case, and is consistent with values that have been employed in other studies of massive stars \citep[e.g.,][]{XuandLi2004a,XuandLi2004b}.

    \begin{table}[ht]
    \centering
    \caption{Overshooting parameters ($f$, $f_0$) applied to the hydrogen-burning core and non-burning shell regions.}
    \label{tab:overshoot}
    \begin{tabular}{lcc}
    \hline
    \textbf{Region} & \textbf{$\mathbf{f}$} & \textbf{$\mathbf{f_0}$} \\
    \hline
    \multicolumn{3}{c}{\textbf{Set 0 — No Overshoot}} \\[1ex]
    \hline
    Hydrogen-burning core & - & - \\
    Non-burning shell      & - & - \\[1ex]
    \hline
    \multicolumn{3}{c}{\textbf{Set 1 — Low Overshoot}} \\[1ex]
    \hline
    Hydrogen-burning core & 0.012 & 0.002 \\
    Non-burning shell      & 0.022 & 0.002 \\[1ex]
    \hline
    \multicolumn{3}{c}{\textbf{Set 2 — Intermediate Overshoot}} \\[1ex]
    \hline
    Hydrogen-burning core & 0.020 & 0.004 \\
    Non-burning shell      & 0.030 & 0.004 \\
    \hline
    \multicolumn{3}{c}{\textbf{Set 3 — High Overhoot}} \\
    \hline
    Hydrogen-burning core & 0.030 & 0.005 \\
    Non-burning shell      & 0.040 & 0.005 \\
    \hline
    \end{tabular}
    \end{table}

   \subsection{Nuclear Reaction Network}

    Nuclear reactions are responsible for energy generation and the chemical evolution of the star. 
    Considering moderate- and high mass stars, we chose the comprehensive \texttt{'pp\_and\_cno\_extras\_o18\_ne22'} network, which comes from the Nuclear Astrophysics Compilation of Reaction rates (NACRE) and the Joint Institute for Nuclear Astrophysics REACtion LIBary (JINA REACLIB) database \citep{Angulo1999, Cyburt2010}. 
    It tracks 19 specific isotopes from the the proton-proton (pp) chain, and the Carbon-Nitrogen-Oxygen (CNO) cycle (including hot CNO reactions), as well as heavier element burning reactions: $^1$H, $^2$H, $^3$He, $^4$He, $^7$Li, $^7$Be, $^8$B, $^{12}$C, $^{14}$N, $^{14}$O, $^{16}$O, $^{18}$O, $^{19}$F, $^{18}$Ne, $^{19}$Ne, $^{20}$Ne, $^{22}$Ne, $^{22}$Mg, and $^{24}$Mg.

    \subsection{Constraints on Numerical Stability}

    The overall spatial and temporal resolution can be managed by scaling \texttt{MESA}'s default settings by the \texttt{mesh\_delta\_coeff} and \texttt{time\_delta\_coeff} {multipliers \citep[see, e.g.,][]{LianJoyce2025}. 
    These values are multipliers on MESA's internal AMR (adaptive mesh refinement)
    In this study, we applied \texttt{MESA}'s default structural resolution by \texttt{mesh\_delta\_coeff = 1.0} and \texttt{time\_delta\_coeff = 1.0}.
    
    The adaptive timestep was regulated by limiting the relative variation of stellar parameters (e.g., $\log T_{\text{eff}}$, $\log g$, $\log L$, $\log T$, $\log \rho$) between each step. 
    This control, specified by \texttt{varcontrol\_target = 1d-4}, helps mitigating the effects of sudden evolutionary changes, such as the main-sequence turn-off, or the core He ignition phase.
    Furthermore, rapid changes in a star's position on the Hertzsprung-Russell (HR) diagram between each step were also controlled by \texttt{delta\_HR\_limit = 0.002}. 

    To ensure enhanced numerical stability and an improved energy conservation, the energy equation of the star was used by the \texttt{dedt} option (\texttt{energy\_eqn\_option = 'dedt'}). Additionally, the \texttt{MESA} standard '\texttt{gold2}' tolerance was applied for tightened numerical tolerances, guaranteeing a stringent numerical criterion and a high numerical accuracy on our stellar models \cite[see more details in][]{Paxton2018, Paxton2019}.

    Furthermore, to prevent numerical instabilities on the convective boundaries, such as convective core splitting, \texttt{MESA}'s  predictive mixing features were utilized by the \texttt{predictive\_superad\_thresh = 0.01} and \texttt{predictive\_avoid\_reversal = 'he4'}.
    Additionally, the \texttt{make\_gradr\_sticky\_in\_solver\_iters} parameter was enabled to further enhance numerical stability during iterations on the boundaries of convective and radiative zones.

\section{Asteroseismic Analysis}

\begin{figure}
    \centering
    \includegraphics[width=\columnwidth]{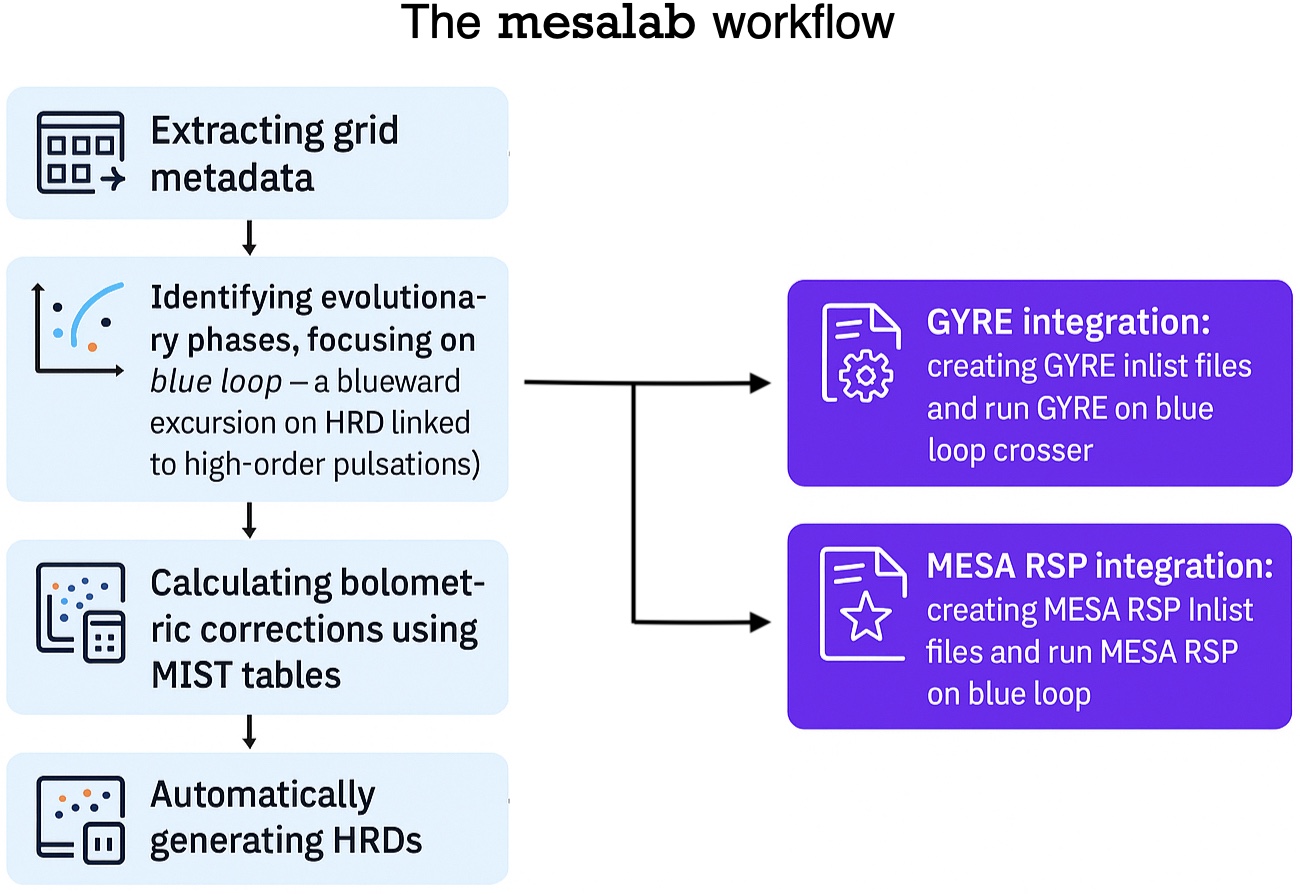}
    \caption{Schematic view of the data processing and visualization of the \texttt{mesalab} pipeline.}
    \label{fig:mesalab_flowchart}
\end{figure}

    To perform the asteroseismic analysis, we employed the stable 7.0 version of the \texttt{GYRE} code \citep{Townsend2013,Townsendetal2018,GoldsteinandTownsend2020}. 
    In a similar work focusing on the case of Mira star R Hya \citep{Joyceetal2024}, the method 'gyre on-the-fly' \citep{Townsend2019MSC,Bellinger2022} was employed for asteroseismic analysis, which allows the user to achieve data during the stellar evolution calculations. In contrast, in the current study, we constructed a pipeline called \texttt{mesalab} by \cite{TarczayNehezd2025} for a more versatile data post-processing method.

\subsection{The \texttt{mesalab} Pipeline}
\label{sec:mesalab}
    
    The \texttt{mesalab} pipeline was developed specifically for this study to handle post-processing and to execute asteroseismic simulations across the full model grid automatically.
    To avoid unnecessary computation, \texttt{mesalab} automatically identifies models exhibiting the blue loop phase and runs \texttt{GYRE} and \texttt{MESA RSP} only on this pre-filtered subset of models.
    These runs operate on the structural models output by \texttt{MESA} at each timestep, which serve as the initial conditions for the oscillation codes.
    
    Figure \ref{fig:mesalab_flowchart} presents a schematic view of the \texttt{mesalab} pipeline workflow. The  pipeline processes the evolutionary grid data produced by \texttt{MESA} through multi-step stages:

    \begin{enumerate}
        \item First, it selects evolutionary tracks that cross the blue loop, based on their trajectories within the Hertzsprung-Russell diagram. 
        This allows us to save computational effort and time by isolating the relevant data.
        
        \item In the next step, the pipeline identifies the specific models that correspond to the blue loop phase. 
        Specifically, it detects the snapshots when the star enters the instability strip from the cooler (red giant branch) side, and the subsequent return point beyond the red edge. 
        Based on these boundaries, the total duration of the blue loop is computed for each model.

        \item In the third step, the pipeline generates \texttt{GYRE} and \texttt{MESA RSP} \citep{Paxton2019} inlist files for the previously selected steps (model numbers), based on a user-specified inlist template.

        \item In the final step, the pipeline executes either \texttt{GYRE} or \texttt{MESA RSP}, depending on the user's specification in the YAML configuration file. 
        These modules are run sequentially, and the number of parallel processes allocated to each can be adjusted by the user to optimize computational efficiency for the given hardware setup.

    \end{enumerate}

    Besides the automated Cepheid blue loop crossing filtering and the sequential execution of \texttt{GYRE} and \texttt{MESA RSP} modules, the pipeline also offers a generic visualization tool.
    It can plot separate stellar evolutionary tracks for each of the grid, generate Colour-Magnitude Diagrams (CMDs) of filtered blue loop candidates from the mesa Isochrones \& Stellar Tracks (MIST) Bolometric Correction Tables\footnote{\url{https://waps.cfa.harvard.edu/MIST/index.html}} \citep{MIST0,MIST}, and plot a heatmap providing a general view of the number of Instability Strip boundary crossings during the blue loop phase.
    The source code of \texttt{mesalab} is publicly hosted at the GitHub repository of the Konkoly Seismolab group\footnote{\url{https://github.com/konkolyseismolab/mesalab}, stable release v1.0.0} and is also available via PyPI. The pipeline remains under active development, as we plan to extend its functionality beyond the present study to support broader applications. A detailed description of the workflow and usage is provided in the official documentation\footnote{\url{https://mesalab.readthedocs.io/en/latest/}}.

    \subsection{Pulsation Stability}
    \label{sec:puls_stability}

 To compute non-adiabatic modes, we assume heat transfer between pulsating and surrounding gas elements (energy transfer). 
  The real part of the work integral $Re(W)$ measures this energy exchange through thermodynamic processes of the pulsation during one cycle. 
  In case of positive work integral ($Re(W)>0$), energy is transferred to the oscillation, leading to an amplified pulsational instability, causing its amplitude to grow.
  To the contrary, in the case of  a negative work integral ($Re(W)<0$), energy is drained from the pulsation, leading to damping. 
  The pulsational instability (growth or damping) is characterized by the $\eta'$ normalized growth rate, as defined by \citet{Stellingwerf-1978} and is described in the \texttt{GYRE} documentation\footnote{\url{https://gyre.readthedocs.io/}}. The normalized growth rate is calculated as the ratio of the positive and negative areas under the work integral curve:
  \begin{equation}
      \eta' = \frac{\int (\oint P dV)dm}{\int |(\oint P dV)|dm}
  \end{equation}
  
This parameter is proportional to the net work done on the pulsation and is commonly used as a measure of pulsation stability. Since the normalized growth rate is used as \texttt{eta} in \texttt{MESA} and \texttt{GYRE}, and as $\eta$ in many publications, we follow that convention here, as well.
  
   \begin{itemize}
       \item $\eta>0$: the pulsation mode is unstable (amplified, or driven), meaning that the amplitude of the pulsation will grow with time, with $\eta=1.0$ meaning that the entire envelope drives;
       \item  $\eta<0$: the pulsation mode is stable (damped), indicating that the amplitude of the pulsation will decay over time, with $\eta=-1.0$ meaning that the entire envelope is dissipative;
       \item $\eta \simeq 0$: the pulsation mode is neutrally stable, indicating that the amplitude of the pulsation neither grows nor decays significantly over time. 
   \end{itemize}

    As a first-order approximation, we performed linear non-adiabatic stability analysis using \texttt{GYRE} and \texttt{MESA RSP}, computing the complex eigenfrequencies and corresponding growth rates ($\eta$) for radial modes. 
    The presence of unstable modes ($\eta > 0$) was used to map the instability domains across the pre-filtered model grid.

\subsection{\texttt{GYRE} and \texttt{MESA RSP} Input Parameters}

\label{sec:gyre_mesa_input}

    To investigate the pulsation behavior of strange mode Cepheid candidates, we initialized \texttt{GYRE} with a range of specific parameters. 
    We mostly followed the default \texttt{GYRE} parameters, with a few minor adjustments, focusing only on the lowest 15 radial modes ($\ell=0$, $m = 0$, $n = 0 - 14$), where strange modes are theoretically expected to emerge \cite[see, e.g.,][]{Buchleretal1997,BuchlerandKollath2001}. 
    The frequency scan was performed as a linear and an inverse scan between 0.01 and 1.0 cycles per day.
    The output data was set to include parameters such as the growth rate ($\eta$) and radial displacement ($\xi_r$), which were critical for confirming the excitation region of the mode to identify strange modes.
    
    We tested multiple solver configurations, including 4$^{th}$ and 6$^{th}$ order Gauss–Legendre Magnus method (\texttt{MAGNUS\_GL4} and \texttt{MAGNUS\_GL6}) \citep{Townsend2013,Townsendetal2018,GoldsteinandTownsend2020}. 
    The latter has been shown to perform more reliably in evolved stellar models, as noted by \citet{Joyceetal2024, LianJoyce2025, Townsendetal2025}, who found it more effective than the default second-order \texttt{diff\_scheme} in resolving narrow instability features.
    
    The calculations were performed using both adiabatic and non-adiabatic mode search routines. 
    As per \texttt{GYRE}'s internal workflow, the solver first performs an adiabatic search to locate approximate eigenfrequencies, which are then used as initial guesses for the non-adiabatic calculations. 
    The output parameters included the growth rate ($\eta$), radial displacement ($\xi_\mathrm{r}$) eigenfunction. 
    These quantities were used to characterize the internal structure of the star and to identify regions of potential instability. 

\begin{figure*}
    \centering
    \includegraphics[width=\textwidth]{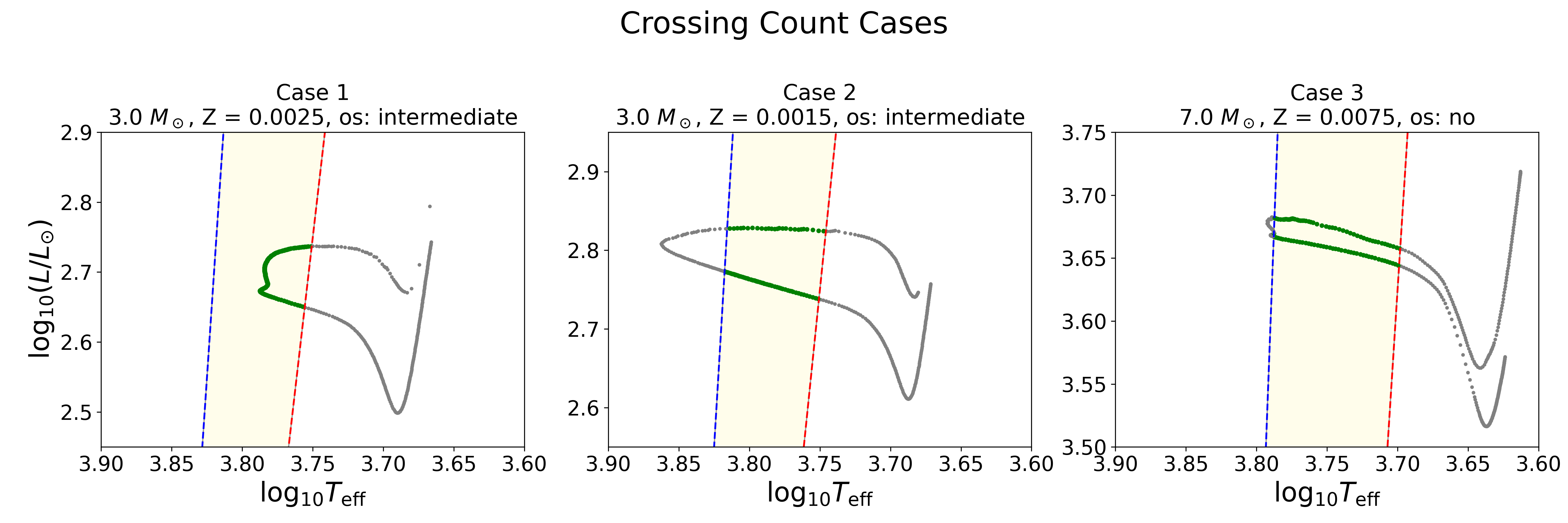}
    \caption{3 cases of instability strip crossing over the blue loop phase. Case 1: the star enters from the red edge and turns back ($C_\mathrm{cross} = 1$, left panel). Case 2: the star passes through both red and blue edges and returns to the AGB phase ($C_\mathrm{cross} = 2$, middle panel). Case 3: the star crosses both edges of the instability strip, briefly shifts redward beyond the blue edge, then re-enters from the blue side before finally returning through both edges toward the AGB phase ($C_\mathrm{cross} = 3$, right panel).}
    \label{fig:crosshrd}
\end{figure*}

    To complement the \texttt{GYRE} analysis, we performed a linear \texttt{MESA RSP} run using a static model configuration. 
    The physical parameters - mass, effective temperature, luminosity, and metallicity - were extracted directly from the \texttt{history.data} and \texttt{inlist} files of the selected \texttt{MESA} model by \texttt{mesalab}. 
    The \texttt{RSP} setup was based on the \texttt{rsp\_Cepheid} test case, which includes a non-linear time evolution phase. However, in our implementation, we retained only the initial linear non-adiabatic stability analysis and disabled further time evolution. The goal was not to simulate full pulsation cycles, but to identify high-order radial excitations. As in the \texttt{GYRE} setup, the configuration was set to search for the first 15 radial modes ($\ell = m = 0$, $n = 0 - 15$).

\subsection{Estimating the High-Overtone Phase, Blue Loop Duration Fraction}

\label{sec:mathcalp}
    
    After identifying the subset of models exhibiting strange mode pulsations, we examined the duration of this phase within the blue loop. From an observational standpoint, it is important to estimate how long a star remains in a state where such pulsations are physically allowed and potentially detectable. To quantify this, we introduce a dimensionless variable, $\mathcal{P_\mathrm{s}}$, defined as the potential of strange mode pulsation:

    \begin{equation} 
    \label{eq:ppot}
        \mathcal{P_\mathrm{s}} = \frac{\tau_\mathrm{s}}{\tau_\mathbf{IS}}. 
    \end{equation}
    
    \noindent Here, $\tau_\mathrm{s}$ is the time (in years) the star spends in the high-overtone strange mode unstable regime, as determined from \texttt{RSP} models.
    Additionally, $\tau_\mathrm{IS}$ denotes the cumulative duration spent within the instability strip, including all intervals between entry at the red edge and exit at the blue edge, accounting for multiple re-entries if present.
 
    Values of $\mathcal{P_\mathrm{s}}$ close to unity indicate that strange modes last longer than the instability crossing phase throughout most of the blue loop, enhancing their observability.
    Conversely, lower values suggest that strange modes appear only during a short period of their lifetime on the blue loop compared to the instability crossing phase, reducing their detectability.

    \subsection{A Metric for Blue Loop Crossing Count: $C_\mathrm{cross}$}

    \label{sec:cross_count_metr}
    
    In this paper, the instability strip was defined using canonical boundaries in the Hertzsprung–Russell diagram, broadly consistent with values adopted in \texttt{MESA} tutorials\footnote{See e.g. Radek Smolec’s tutorial in the \href{https://mesastar.org/summer-school-2023/}{MESA Summer School 2023}.}. 
    This definition allows us to identify boundary crossings during each stellar evolution across the model grid. 
    The crossing count, $C_\mathrm{cross}$, indicates how many times the stellar track intersects back and forth a boundary of the instability strip (see Figure\,\ref{fig:crosshrd}). 
    For example:
    
    \begin{itemize}
        \item $C_\mathrm{cross} = 0$: indicates that the star does not cross the instability strip red edge during its evolution, i.e., it never enters the strip from the red side and does not perform a blue loop.
        \item $C_\mathrm{cross} = 1$: corresponds to a single entry from the cool side (red edge) into the IS, followed by an exit back to the AGB phase (without exiting through the blue edge).
        \item $C_\mathrm{cross} = 2$: indicates entry from the red edge, exit through the blue edge, and a return toward the asymptotic giant branch (AGB).
        \item $C_\mathrm{cross} = 3$: represents a more complex path in which we observe a kink around the blue edge: entry via the red edge, exit through the blue edge, re-entry from the blue side, a second exit through the blue edge, and a final return through both blue and red edges toward the AGB.
    \end{itemize}

\section{Results}    

    In the following sections, we present our results on the impact of overshoot prescriptions and the physical properties of the star (mass and metallicity) on the explored parameter space.

\begin{figure*}
    \centering
    \includegraphics[width=\textwidth]{fig3.png}
    \caption{The number of blue loop crossings in each parameter set is shown on the mass–metallicity plane. Set 0 corresponds to the reference models without convective overshoot, while Sets 1–3 represent increasing overshoot efficiency. It can be clearly seen that higher overshoot suppresses blue loop excursions at the upper mass range. Moreover, in Set 0, the blue loop crossing regime fragments into two islands: at intermediate metallicities, blue loop crossings are almost absent across the entire mass range. Note that each crossing event corresponds to a single edge of the instability strip, traversed in both directions. For example, a crossing count of $C_\mathrm{cross} = 1$ indicates that the star entered and exited through the red edge only, without reaching the blue boundary (see more details in the text).}
    \label{fig:heatmaps}
\end{figure*}

\subsection{Tracking Instability Strip Crossings During the Blue Loop Phase
}

    After filtering the blue loop crossing with \texttt{mesalab}, the number of relevant models was reduced from 1026 to 438, 529, 440, and 311 for Sets 0, 1, 2, and 3, respectively.
    As defined in Section\,\ref{sec:cross_count_metr}, the $C_\mathrm{cross}$ metric allows us to quantify how many times a given star crosses the instability strip boundaries—blue and/or red—during its blue loop phase. 
    Each full crossing, i.e., entry and exit through the same boundary, is counted as one.
    Figure\,\ref{fig:heatmaps} gives a comprehensive view of the $C_\mathrm{cross}$ on the $M-Z$ plane for each overshoot case (Sets 0 - 3). 
    The most striking trend is the continuous narrowing of the blue loop crossing region as overshoot increases. 
    In the following, we examine each model set individually to quantify how overshoot affects the extent and frequency of instability strip crossings.

\subsubsection{Set 0: The Canonical Model}
\label{sec:set0_blueloops}

    In the control group (Set 0, without convective overshoot), the most prominent structural feature is the 
    splitting of the blue loop regime into two distinct islands on the mass–metallicity plane (see Set 0 panel of Figure\,\ref{fig:heatmaps}).
    The overall blue loop crossing regime spans from $2.5$ to $12.5\, M_\odot$. 
    The lower mass boundary tends to increase with metallicity, while the upper boundary remains nearly constant around $12 - 12.5\, M_\odot$. 
    
    The separation of the two islands appears between $Z = 0.0080 - 0.0100$. 
    We found that, in the case of low-metallicity models, the lower mass threshold shifts upward with increasing $Z$, while the outer boundary contracts, peaking at $Z = 0.0075$ and $M = 6.5 - 7.5\, M_\odot$. 
    The higher metallicity island starts at $Z = 0.0105$ with masses of  $8.5 - 10\, M_\odot$, then the regime expands as $Z$ increases.
    Namely, the lower mass threshold drops down to $4.5 - 5\, M_\odot$, while the high mass threshold increases to $12.5\,M_\odot$. 
    
    We found some cases, when the value of the crossing count reaches three (e.g., for $Z = 0.0075$, $M=7.0\,M_\odot$), indicating multiple transitions across the instability strip boundaries during the blue loop phase, as illustrated in the right panel of Figure\,\ref{fig:crosshrd}.

\subsubsection{Set 1: Low overshoot}

    In the low overshoot setup (Set 1), instability strip crossings  $C_\mathrm{cross}\ge 1$ occur mainly in intermediate- and high-mass models.
    We found that as metallicity increases, the lower mass threshold rises as well (see panel Set 1 of Figure\,\ref{fig:heatmaps}).
    For low metallicities ($Z\le 0.0035$), blue loop crossings occur already at 
    $M\ge 3.0\,M_\odot$, and the mass range extends up to $\sim11.0\,M_\odot$.
    
    In the intermediate metallicity range ($0.0040\le Z \le 0.0100$), the crossings appear toward higher masses, typically starting around 
    $4.0 - 5.0\,M_\odot$, while the upper mass limit remains near $10.0 - 10.5\,M_\odot$.
    For high metallicities ($Z\ge0.0105$), crossings are shifted to more massive models ($M\ge 6.0\,M_\odot$), with the lower bound increasing steadily. 
    At the highest metallicities ($Z\ge 0.0185$), crossings persist up to the maximum mass considered ($11.5\,M_\odot$).
    A similar trend is observed in the second dataset, where the mass range associated with $C_\mathrm{cross}\ge1$ shifts toward higher masses with increasing $Z$.

    \subsubsection{Set 2: Intermediate overshoot}

    In the intermediate overshoot setup (Set 2), instability strip crossings with a count of $\ge 1$ again fall to intermediate- and high-mass models, but the corresponding mass range shifts upward and becomes somewhat narrower compared to Set 1  (see panels Set 1 and 2 of Figure\,\ref{fig:heatmaps}).
    As metallicity increases, the lower mass limit for such crossings rises systematically.
    For low metallicities ($Z\le 0.0030$), blue loop crossings occur at $M\ge 3.0\,M_\odot$, with the upper bound extending to 
    $10.5\,M_\odot$. 
    In this regime, the mass range is relatively broad, occasionally split into multiple intervals (e.g., $M = 3.0 - 3.5$ and $\sim 5.0 – 10.5\,M_\odot$ at 
    $Z = 0.0025$).
    In the intermediate metallicity range ($0.0040\le Z \le 0.0105$), the lower boundary shifts to $M \approx 4.0 - 5.0$, while the upper limit remains near $\approx 9.0 - 10.5\, M_\odot$.
    The mass intervals become narrower and more continuous, with fewer low-mass entries.
    At high metallicities ($Z \ge 0.0110$), crossings are restricted to increasingly massive models, starting at 
    $M\ge 5.5 \,M_\odot$ and extending up to 
    $10.5 \,M_\odot$. 
    The lower bound continues to rise with 
    $Z$, reaching 
    $M\ge 6.5\,M_\odot$ at the highest metallicities ($Z = 0.0185 – 0.0200$).

    Overall, the trend observed in Set 2 confirms that  
    $C_\mathrm{cross}\ge 1$ occurs within a slightly narrower and higher mass range as metallicity increases. 
    This behaviour reflects the combined influence of metallicity on stellar structure, envelope response, and the extent of blue loop excursions during core helium burning.

    \subsubsection{Set 3: High Overshoot}

    In the high-convection configuration (Set 3), instability strip crossings with a count of $\ge 1$ occur across a wide range of intermediate- and high-mass models, but the mass intervals shift upward with increasing metallicity and gradually narrow toward higher $Z$ compared to Sets 1 and 2  (see panel Set 3 of Figure\,\ref{fig:heatmaps}).
    At low metallicities ($Z\le 0.0025$), crossings appear already at $M\ge 3.0\,M_\odot$, with the broadest mass range observed at $Z=0.0015$. 
    As mass increases, additional metallicity values begin to show crossing behavior, including split intervals (e.g., $M=3.5$ and $7.0-9.0\,M_\odot$ at $Z=0.0025$).
    In the intermediate metallicity regime $(0.0030\le Z\le 0.0100)$, the lower mass boundary rises gradually from $M \simeq 4.0\,M_\odot$ to $5.5\,M_\odot$, while crossings continue to occur up to $M\simeq 9.0 - 9.5\,M_\odot$.
    The mass intervals become more continuous and shift toward higher values.
    At high metallicities ($Z\ge 0.0105$), crossings are restricted to increasingly massive models, starting at $M\ge 6.0\,M_\odot$ and extending to $9.0-9.5\,M_\odot$. 
    The lower boundary continues to rise with $Z$, reaching $M\ge 8.0\,M_\odot$ at the highest metallicity ($Z=0.0200$).

    Overall, the mass range associated with $C_\mathrm{cross}\ge 1$ shifts toward higher masses as metallicity increases, and becomes narrower at the low-mass end.

\subsection{HRD Morphology and the CMD Projection of Blue Loop Models}

\begin{figure*}
    \centering
    \includegraphics[width=\textwidth]{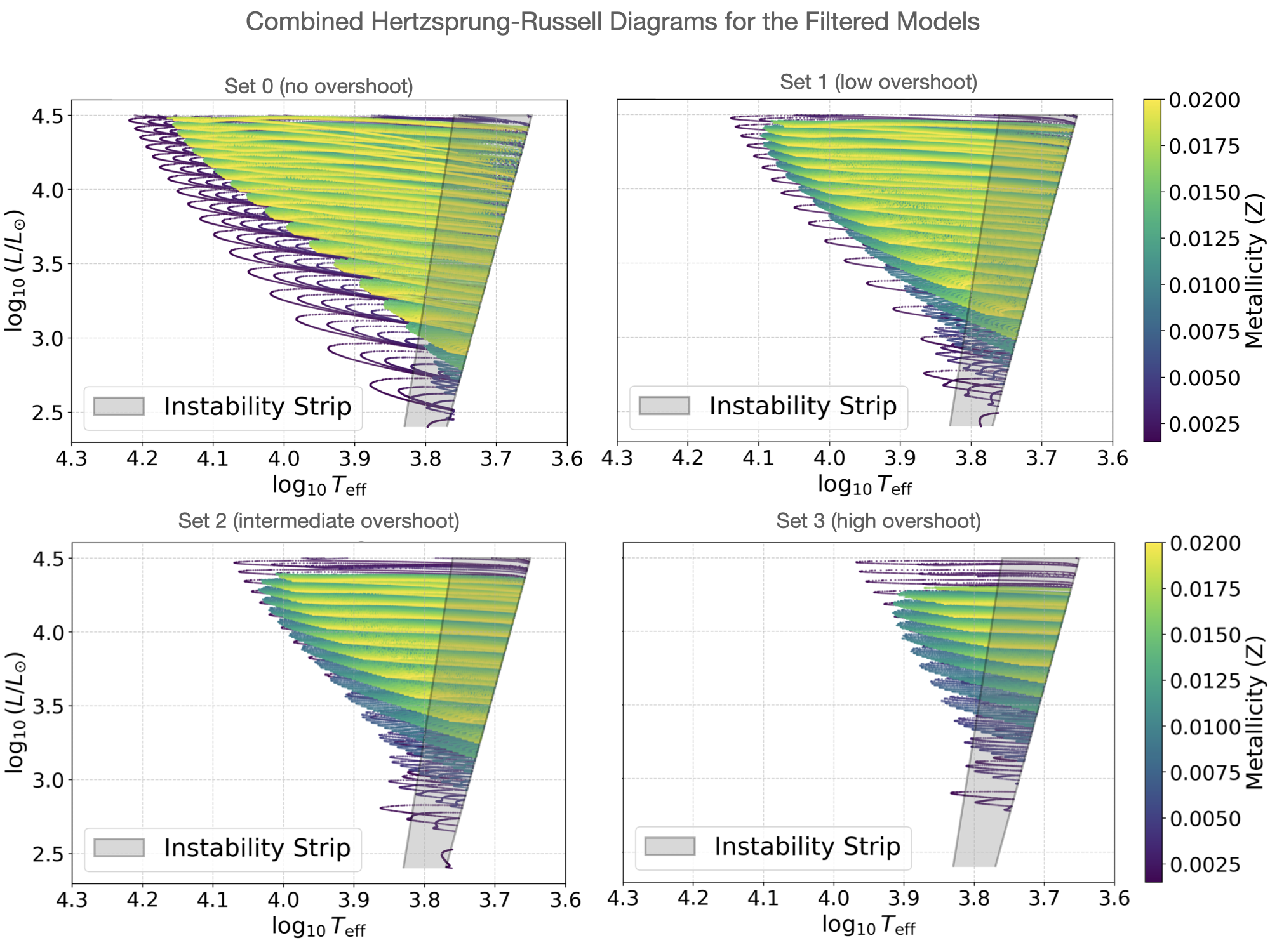}
    \caption{Hertzsprung–Russell diagram showing all selected models that undergo a blue loop phase and cross the instability strip from the red edge for all the overshoot configurations (Sets\,0-3) color-coded by the $Z$ metallicity. It is clearly visible that the extent of the loop increases with decreasing metallicity and also with increasing luminosity. Furthermore, a clear dependence on overshoot is also present. As the overshoot parameter increases, the number of models having a blue loop phase decreases significantly. These models were used as input for \texttt{GYRE} and \texttt{MESA RSP} asteroseismic simulations. The extension of the instability strip is adopted from canonical \texttt{MESA} tutorials. }
    \label{fig:combined_hrd_filt}
\end{figure*}

    As noted in Section\,\ref{sec:mesalab}, \texttt{mesalab} generates a composite HR diagram and a \textit{Gaia} color–magnitude diagram for the filtered blue loop models. 
    To compute synthetic magnitudes in the \textit{Gaia} bands, \texttt{mesalab} utilizes the \texttt{isochrones} Python package, which provides access to the MIST bolometric correction tables \citep{MIST0, MIST}. 
    This step is essential to be able to place the potential strange mode Cepheids on both the $\log T_\mathrm{eff} - \log L$ and the  $M_\mathrm{G}$ vs. $M_\mathrm{BP} - M_\mathrm{RP}$ planes.

    Figure\,\ref{fig:combined_hrd_filt} and the Figure\,\ref{fig:combined_cmd_filt} in Appendix\,\ref{sec:appendixHRD} show the distribution of blue loop crossing models in both the HRD and the \textit{Gaia} color-magnitude diagrams.
    In each figure, the metallicity is color-coded. 
    It is clearly visible in Figure\,\ref{fig:combined_hrd_filt} that the morphology of the blue loop on the HRD is strongly influenced by metallicity.
    Namely, lower-$Z$ models tend to exhibit more extended loops toward higher effective temperatures, independent of stellar mass. 
    Thus, as metallicity increases, the extent of the loop toward the blue becomes more limited.
    
    Moreover, it is also evident in Figure\,\ref{fig:combined_hrd_filt} that stellar mass also plays a significant role: as luminosity increases along the vertical axis, the loop extent grows accordingly, reflecting the mass dependence of the envelope structure and core evolution.
    Furthermore, the figure also reveals a clear trend with overshoot strength: as the overshoot parameter increases (from Set 1 to Set 3), blue loop crossers take shorter excursions in effective temperature. 
    Thus, the blue loop crossing domain itself shrinks: while in Sets 0 and 1 the most extended loops reach $\log T_\mathrm{eff}\simeq 4.1-4.2$, in the case of Set 3, even the longest loops fail to exceed $\log T_\mathrm{eff} = 4.0$. 
    In other words, the blue tip of the loop visibly shifts towards lower temperatures as overshoot increases.
    
    A similar contraction is seen in luminosity: the lower end of the blue loop population, around $\log L/L_\mathrm{\odot} =2.5–2.7$, disappears entirely in Set 3. 
    This indicates that stronger overshoot not only limits the horizontal extent of the loop but also suppresses its emergence in lower-luminosity models. The color-magnitude diagrams preserve the above-mentioned trends (see Figure\,\ref{fig:combined_cmd_filt} in Appendix\,\ref{sec:appendixHRD}). 

\subsection{Asteroseismological Filtering and Mode Selection}

\begin{figure*}
    \centering
    \includegraphics[width=0.48\linewidth]{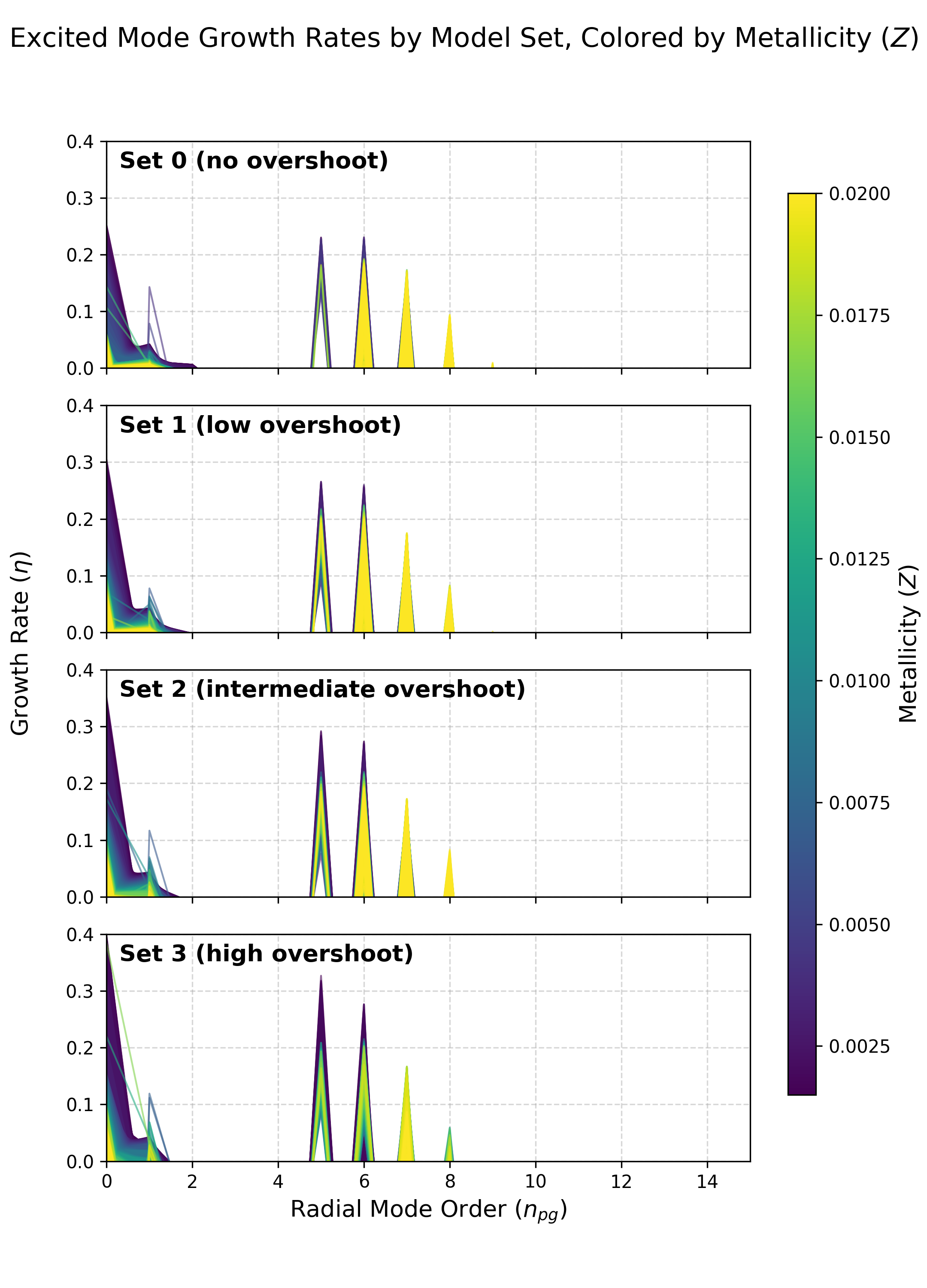}
    \includegraphics[width=0.48\linewidth]{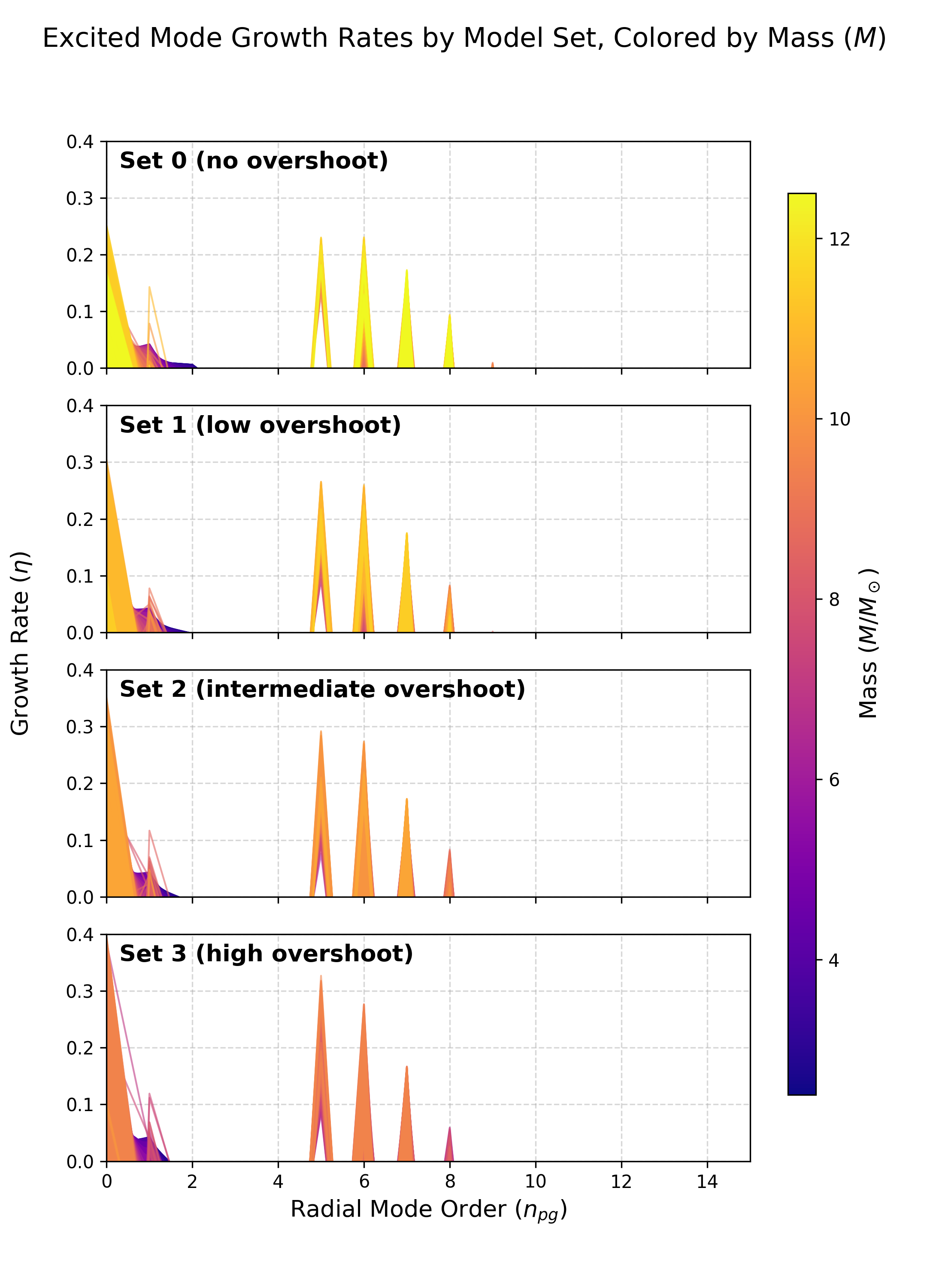}
    \caption{The bimodal nature of pulsational behaviour. All of the models fall into two groups: either dominated by low-order pulsations ($n_\mathrm{pg} = 0, 1,2$ or high-order pulsations $n_\mathrm{pg}\ge5$. The left-hand side panel presents the effect of metallicity on excited pulsation modes, while the right-hand side panel is color-coded by stellar mass for model Sets 0-3. Note that in both panels, curves are plotted in ascending order of metallicity or mass, respectively. This means that at higher radial orders, models with larger $Z$ or $M$ tend to visually obscure those with lower values. However, the overplot is less pronounced at lower radial orders, allowing the effect of low-metallicity or low-mass models to be visible more clearly - particularly in those models, where their growth rates are slightly higher. For a complementary view without overplot, see Figure\,\ref{fig:eta_2d_heatmap}, which shows 2D heatmaps of maximum growth rate across radial order and metallicity (left) or mass (right). }
    \label{fig:pos_etas}
\end{figure*}

\begin{figure*}
    \centering
    \includegraphics[width=0.48\textwidth]{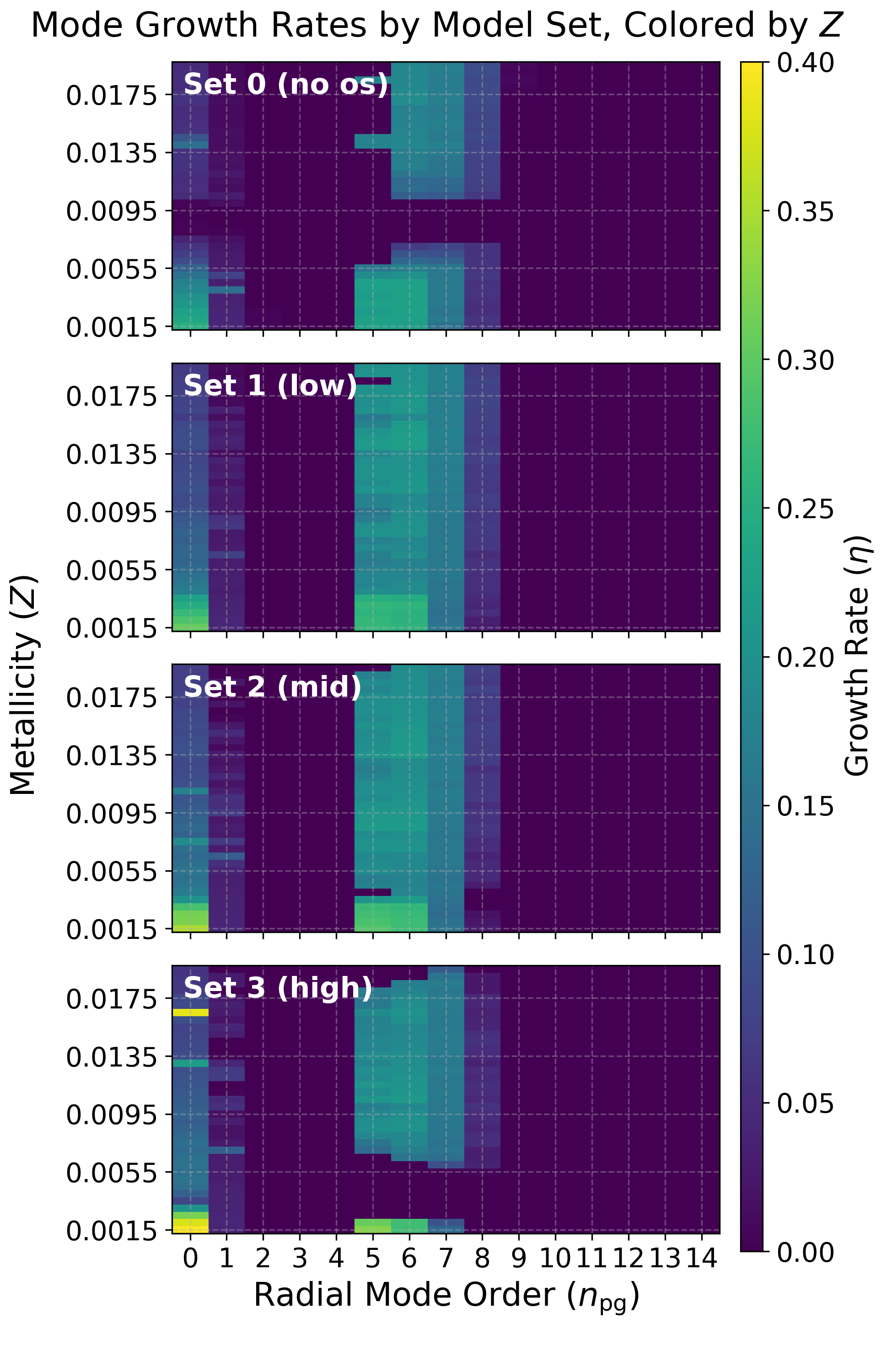}
    \includegraphics[width=0.48\textwidth]{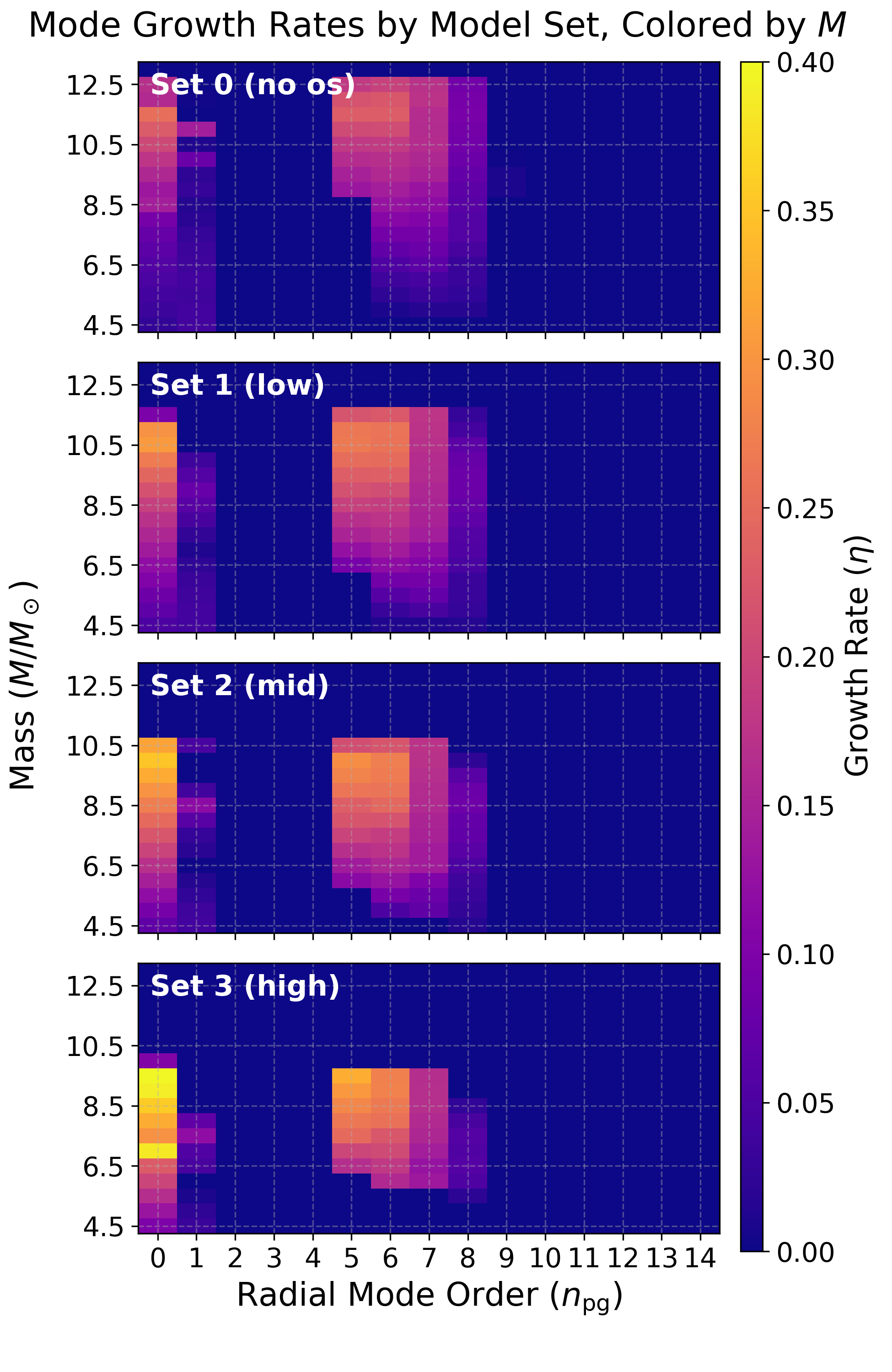}
    \caption{Growth-rate ($\eta$) maps in the $n_\mathrm{pg}$–$Z$ (right) and $n_\mathrm{pg}$–$M$ (left) planes, shown for each overshoot configuration (Sets\,0–3).
    Both panels present models with positive growth rates ($\eta > 0$) only.  
    Color indicates the value of $\eta$, with brighter regions corresponding to stronger excitation.  
    Fundamental and low-order overtone modes ($n_\mathrm{pg} = 0$–$2$) appear to be separated from the strange modes ($n_\mathrm{pg} \geq 5$).
    There is no apparent transition between these two excitation domains, suggesting that strange modes emerge independently of the classical mode pulsations.
    A clear correlation is seen between the investigated overshoot setup and the frequency and extent on both planes of mode excitation.  
    As overshoot increases from Set\,0 to Set\,3, the excitation domain in both the $n_\mathrm{pg}$–$M$ and $n_\mathrm{pg}$–$Z$ planes becomes narrower. 
    Namely, strange modes ($n_\mathrm{pg} \geq 5$) shift toward lower and intermediate-mass stars, and their occurrence becomes more restricted to specific metallicity ranges.
}
    \label{fig:eta_2d_heatmap}
\end{figure*}

    As described in Section\,\ref{sec:gyre_mesa_input}, we performed \texttt{GYRE} and \texttt{MESA RSP} analyses on all filtered blue loop crossing models. 
    Despite testing multiple \texttt{GYRE} configurations (e.g., different \texttt{diff\_scheme} solvers and frequency scan ranges; see details in Section\,\ref{sec:gyre_mesa_input}), we did not detect excited modes above the second overtone. 
    For $n_\mathrm{pg}\ge3-4$, $\eta$ was consistently negative, or the computation failed to converge. 
    We consider this to be either a parameter sensitivity or the limitations of \texttt{GYRE}’s linear treatment of mode trapping near a steep potential gradient. 
    In contrast, we successfully identified a model with excited high overtone pulsation modes utilizing the \texttt{MESA RSP} module. 
    In the following, all the asteroseismic results are derived from non-adiabatic linear pulsation analyses performed with the \texttt{MESA RSP} module.

\subsubsection{Isolating Strange Mode Pulsations}
\label{sec:strange_mode_selection}

    Our asteroseismic simulations reveal a bimodal nature in pulsational behavior. 
    Each model falls into one of two categories: either low-order modes are excited (fundamental, first, or second overtone), with no higher-order instability present; or instability appears only in high-overtone modes ($n_\mathrm{pg} \ge 5$), typically without excitation of the fundamental mode (see Figure\,\ref{fig:pos_etas}). 
    No cases were found where both low- and high-order modes were simultaneously unstable, nor did we find a transition of instability near mode orders three and four. 
    Thus, to isolate models with high-overtone instability, we applied a reasonable selection above the threshold of $n_\mathrm{pg}^\mathrm{th} = 3$.
    This way, we excluded those models that are dominated by low-order excitations. 
    All subsequent analysis of strange mode behavior is restricted to models meeting this criterion.

    Note that in both panels of Figure\,\ref{fig:pos_etas}, curves are plotted in ascending order of metallicity or mass, respectively. 
    This means that at higher radial orders, models with larger $Z$ or $M$ tend to visually obscure those with lower values. 
    However, the overplot is less pronounced at lower radial orders, allowing the effect of low-metallicity or low-mass models to be visible more clearly - particularly in those models, where their growth rates are slightly higher.
    For a clearer view of the overall distribution, Figure\,\ref{fig:eta_2d_heatmap} presents the same data as Figure\,\ref{fig:pos_etas} as 2D heatmaps of the maximum growth rate. 
    The values are binned by radial order and either metallicity or mass. 
    This representation avoids overplot and reveals the bimodal structure more pronounced.

    After isolating strange modes by the $n_\mathrm{pg}^{th}$ threshold, we examined their structural properties in detail.
    Appendix\,\ref{app:strange_mode_struct} presents two representative stellar models (Figures\,\ref{fig:eigen_work_app1} and \ref{fig:eigen_work_app2}), where the radial displacement and work integrals are shown for all radial modes. 
    In both cases, the strange mode stands out clearly from the others, either in the case of the fifth (Figure\,\ref{fig:eigen_work_app1}) or the ninth overtone (Figure\,\ref{fig:eigen_work_app2}) pulsation (marked with red lines and $\mathcal{S}$ in the upper left corner). 
    These examples show how strange modes appear as a sharp peak in the work integral (lower panels) in the outermost region of the star, or as a jump in the radial displacement profile (upper panels). 
    These features reflect the surface localized pulsation nature of strange modes.

\begin{figure*}
    \centering    \includegraphics[width=\linewidth]{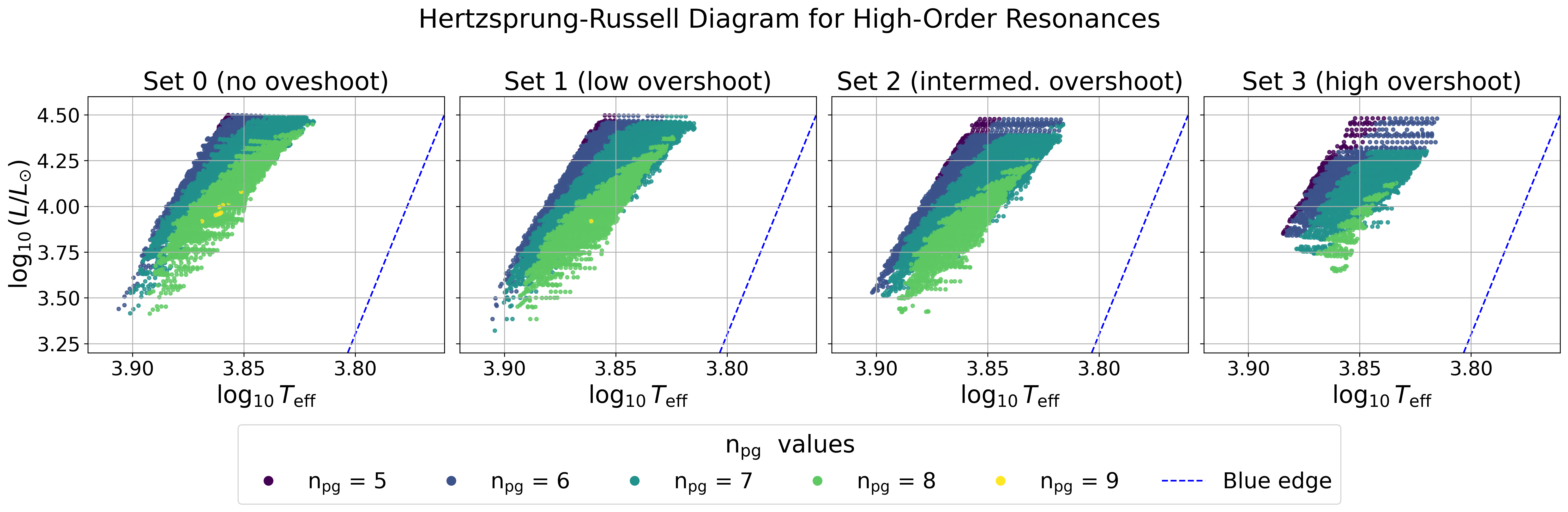}    
    \caption{The HRD for high-order, strange modes for all overshoot sets. Each model is color-coded by the excite mode number ($n_\mathrm{np}$). Considering overshoot prescription, a clear trend is visible: as $\log T_\mathrm{eff}$ and $\log (L/L_\odot)$ decrease, $n_\mathrm{pg}$ decreases. In the case of Sets\,0 and 1, strange mode excitation spans down to $\log(L/L_\odot)\simeq 3.2-3.4$ and $\log T_\mathrm{eff}\simeq 3.90-3.92$. To the contrary, in the high overshoot case (Set\,3), the lower limit of luminosity and temperature shifts to $\log(L/L_\odot\simeq3.65$, and $\log T_\mathrm{eff}\simeq 3.88-3.89$, respectively.
    Furthermore, strange modes with $n_\mathrm{pg} = 5$–$8$ are excited across all overshoot sets, while $n_\mathrm{pg} = 9$ appears only in the no overshoot and low overshoot configurations, and only in a few isolated cases.}
    \label{fig:strange_HRD}
\end{figure*}

\begin{figure*}
    \centering    \includegraphics[width=\linewidth]{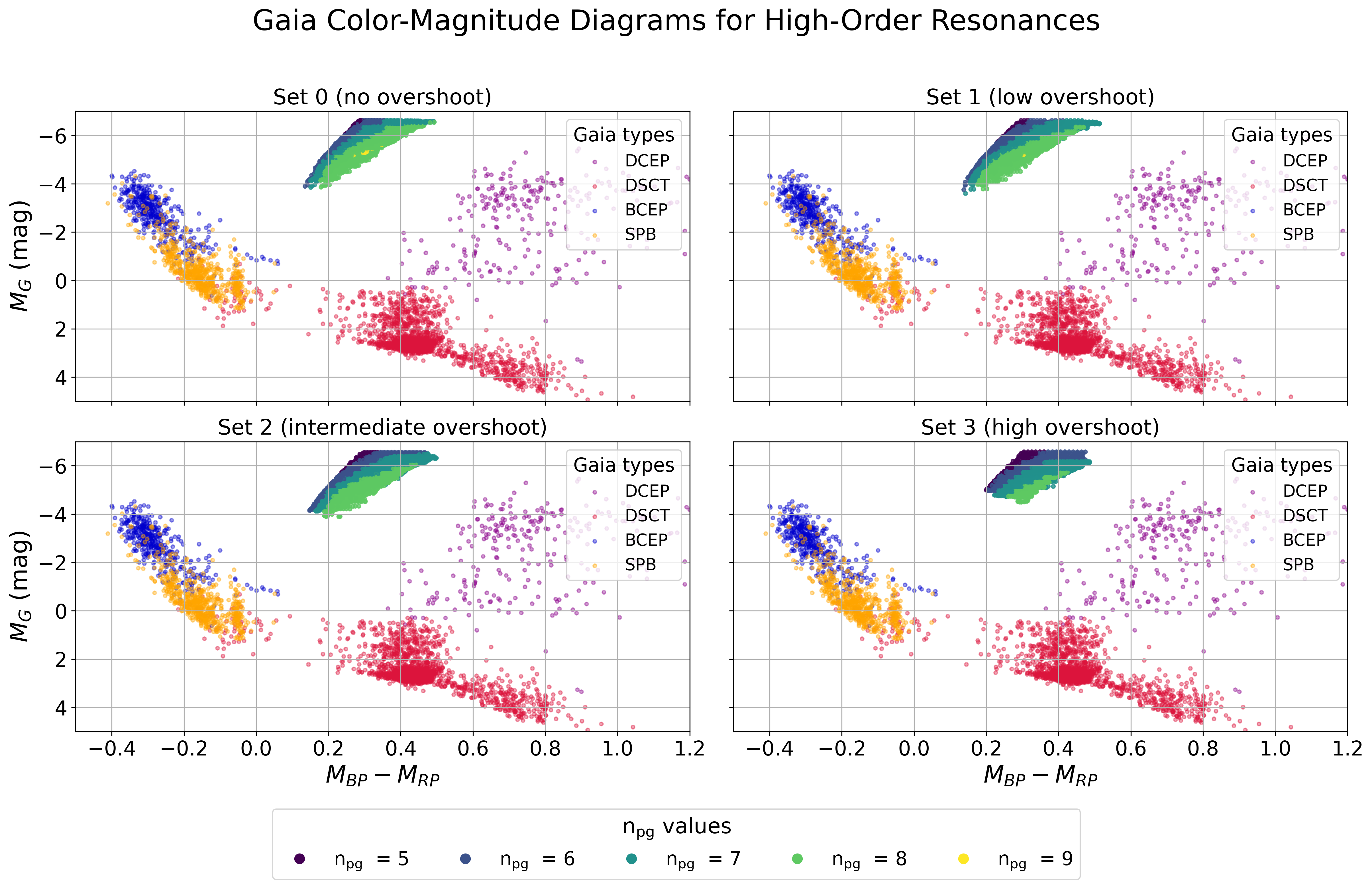}        \caption{Similar to Figure~\ref{fig:strange_HRD}, but projected on the \textit{Gaia} $M_\mathrm{G}$ vs. $(M_\mathrm{BP} - M_\mathrm{RP})$ plane. The diagram is further combined with $\delta$ Scuti and $\gamma$ Doradus stars (DSCT), $\delta$ Cepheids (DCEP), slowly pulsating B stars (SPB), and $\beta$ Cepheids (BCEP) from the \textit{Gaia} DR3 archive, offering a direct comparison between model predictions and observed variability. As in the previous figure, the color-coding refers to the excited mode number ($n_\mathrm{pg}$) for \texttt{MESA RSP} runs. Additionally, \textit{Gaia} variables are color-coded with faint purple, red, blue, and yellow dots, which denote DCEP, DSCT/GDOR, BCEP, and SPB variables, respectively. This projection also reflects the impact of overshoot. As overshoot increases, the region of mode excitation becomes narrower and shifts toward more luminous and redder (cooler) stars with higher color indices.}
    \label{fig:strange_CMD}
\end{figure*}

    To visualize the distribution of strange mode pulsators, we projected all excited models onto the Hertzsprung–Russell diagram (Figure \ref{fig:strange_HRD}) and the \textit{Gaia} $M_\mathrm{G}$ vs.\ $(M_\mathrm{BP}-M_\mathrm{RP})$ color–magnitude diagram (Figure \ref{fig:strange_CMD}).
    
    In both representations, models are color-coded by their excited mode number $n_\mathrm{pg}$, allowing us to trace how mode distribution varies across the temperature–luminosity and color–magnitude spaces.  
    At first glance, a clear trend can be seen: as effective temperature and luminosity decrease, the value of $n_\mathrm{pg}$ increases.  
    Figure\,\ref{fig:strange_CMD} is further extended with $\delta$ Cephei (DCEP), $\delta$ Scuti--$\gamma$ Dor (DSCT/GDOR), $\beta$ Cephei (BCEP), and slowly pulsating B stars (SPB) variable stars (faint purple, red, blue, and yellow background dots) from the \textit{Gaia} DR3 variability surveys \citep{GaiaDR3_varsum_2023,GaiaDR3_varcat_2023,GaiaDR3_Cep_2023}.
    Their inclusion helps to identify the place of strange mode pulsators, compared to the nearby variable star types on the \textit{Gaia} CMD plane.
    BCEP and SPB sources were required to have parallaxes greater than $0.1$ milliarcseconds, \texttt{phot\_g\_n\_obs} $> 40$, and \texttt{parallax\_over\_error} (SNR) $> 20$. For the fainter and more populous DSCT/GDOR sample we increased the parallax limits to 3.5 milliarcseconds and SNR $> 50$, whereas for the DCEP sample, we relaxed the SNR threshold to $> 10$ but also set an extinction limit of \texttt{ag\_gspphot} $<1.5$~magnitudes. In all cases, we used the extinction values found in the DR3 catalog. While these samples still include some systematics (substructures and extensions, especially for the DSCT/GDOR group), they still define where other classes of variables exist relative to the strange mode models.
    
    Notably, the strange mode pulsators form a clearly isolated island in the \textit{Gaia} CMD.
    These stars are shifted towards bluer $(M_\mathrm{BP} - M_\mathrm{RP})$ colours compared to the classical Cepheids, and they show no apparent overlap with any of the \textit{Gaia} DR3 variable types, including DSCT, DCEP, SPB, and BCEP stars. Potential overlaps may occur with some other variables, such with as the PV~Tel class, but those stars are exceedingly rare \citep{pv-tel-2008}.

    The extension of mode excitation on either the $\log T_\mathrm{eff} - \log (L/L_\odot)$ or color-magnitude planes also depends strongly on the adopted overshoot prescription.  
    In the case of no overshoot and Set\,1 configurations, strange mode excitation spans down to $\log(L/L_\odot) \simeq 3.2$–$3.4$ and $\log T_\mathrm{eff} \simeq 3.90$–$3.92$ on the HRD.
    As overshoot prescription increases, the excitation domain shrinks.
    For instance, in Set\,3 models, the lower luminosity limit shifts upward to $\log(L/L_\odot) \simeq 3.65$, and the lower temperature limit shifts to $\log(T_\mathrm{eff}) \simeq 3.88$–$3.89$. 

    This contraction is mirrored in the CMD: the lower $M_\mathrm{G}$ boundary brightens from --3.5 to --4.5, and the $M_\mathrm{BP} - M_\mathrm{RP}$ color index increases from $0.10–0.15$ to $0.20$. 
    
    Despite the wide coverage in mass and metallicity, we did not find any models with excitation of mode $n_\mathrm{pg} = 10-12$, as previously proposed by \citet{Buchleretal1997} and \citet{BuchlerandKollath2001}. 
    The highest excited mode in our grid is $n_\mathrm{pg} = 9$, although this mode appears only in a few no-overshoot and low-overshoot models.

\begin{table*}
\centering
\caption{Summary of excited radial modes ($n_\mathrm{pg}$) across all overshoot Sets.}
\begin{tabular}{ccccccc}
\hline
\textbf{Mode} ($\mathbf{n_\mathrm{pg}}$) & \textbf{Total Count} & \textbf{Dominant Set} & \textbf{Mass Range} ($\mathbf{M_\odot}$) & $\mathbf{Z}$ \textbf{Range} & \textbf{Peak Configuration} & \textbf{Notes} \\
\hline
5 & 752 & Set\,2 & 6–12.5 & 0.0015–0.0200 & $10\,M_\odot$, $Z=0.0015$ & Rare, fragmented distribution \\
6 & 6945 & Set\,2 & 4.5–12.5 & 0.0015–0.0200 & $10.5\,M_\odot$, $Z=0.0030$ &\\
7 & 13\,767 & Set\,1 & 4.5–12.5 & 0.0015–0.0200 & $9.5\,M_\odot$, $Z=0.0015$ & Most frequent mode \\
8 & 7058 & Set\,0 & 4.5–12.5 & 0.0015–0.0200 & $9\,M_\odot$, $Z=0.0200$ & \\
9 & 10 & Set\,0 & 8.5–10 & 0.0175–0.0200 & $9\,M_\odot$, $Z=0.0195$ & Isolated cases only in Sets\,0–1 \\
\hline
\end{tabular}
\label{tab:npg_summary}

\end{table*}

\begin{table*}
\centering
\caption{The distribution of excited radial modes ($n_\mathrm{pg}$) across overshoot Sets. Each cell shows the number of excited models and their fraction relative to the total number of models in the given Set.}
\begin{tabular}{lccccc}
\hline
Set & $n_\mathrm{pg}=5$ & $n_\mathrm{pg}=6$ & $n_\mathrm{pg}=7$ & $n_\mathrm{pg}=8$ & $n_\mathrm{pg}=9$ \\
\hline
Set\,0 & 51 (0.04\%) & 1266 (9.3\%) & 2817 (20.8\%) & 2548 (18.8\%) & 9 (0.007\%) \\
Set\,1 & 209 (2.1\%) & 2301 (16.2\%) & 4660 (32.9\%) & 2479 (17.5\%) & 1 (0.0007\%) \\
Set\,2 & 283 (4.2\%) & 2285 (13.6\%) & 4193 (20.4\%) & 1742 (8.6\%) & 0 (0.0\%) \\
Set\,3 & 209 (0.6\%) & 1042 (2.8\%) & 2097 (5.7\%) & 289 (0.8\%) & 0 (0.0\%) \\
\hline
\end{tabular}
\label{tab:npg_distribution}
\end{table*}

\subsubsection{Statistical Analysis}

    To provide a comprehensive view of the excitation statistics, we present the results in Tables\,\ref{tab:npg_summary} and \ref{tab:npg_distribution}. 
    In the case of the reference model (Set 0), 135\,499 \texttt{MESA RSP} runs were executed on the filtered models, from which 6691 models met the $n_\mathrm{pg} \ge n_\mathrm{pg}^\mathrm{th}$ criterion.
    This means that only $\sim5$\% of the models satisfied the condition.
    For Set\,1, 9650 out of 141\,796 models met the threshold (6.8\%), while 
    Set 2 contained 8503 qualifying models out of 68157 (12.5\%). 
    Set 3 included 3637 out of 36526 models (9.96\%). 
    Even in Set 2, where the fraction is highest, high-overtone instability remains a rare outcome, reflecting the rarity of high-overtone mode excitations.

    To summarize the mass and metallicity distribution among models that satisfy the $n_\mathrm{pg} > n_\mathrm{pg}^{th}$ criterion across the four overshoot sets (Set 0 - 3), we group the results by radial mode order ($n_\mathrm{pg}$), reporting total occurrences, parameter ranges, and dominant values for each mode.

\vskip 1em

\noindent\underline{\textbf{Mode $n_\mathrm{pg} = 5$}}
\vskip 0.5em

    \noindent For mode $n_\mathrm{pg} = 5$, excitation remains rare across all four model sets, with a total of 752 occurrences. 
    Set\,0 contains 51 cases, spanning a mass range of $9 - 12.5\,M_\odot$ and metallicities between $Z = 0.0015$ and $0.0200$.
    The most frequent combination is $11.5\,M_\odot$ at $Z=0.0035$.
    
    We found that Sets 1 and 3 contain 209 excited models for mode $n_\mathrm{pg} = 5$, although their internal distributions differ. 
    Set 1 spans a broader mass range ($6.5-11.5\,M_\odot$), while Set 3 is restricted to a narrower mass range ($6.5 - 9.5\,M_\odot$). Similarly, the metallicity range is also narrower with a higher overshoot envelope.
    While in Set 1 $n_\mathrm{pg} = 5$ mode excitation appears across the whole investigated metallicity range, in the case of Set 3, excitation is restricted to a narrower interval between $Z \simeq 0.0065 - 0.0070$ and $0.0180$.
    Interestingly, the mode also appears at the lowest metallicities tested ($Z = 0.0015 - 0.0020$), indicating that excitation is not strictly confined to high and intermediate $Z$ values (see left-hand side panel of Figure\,\ref{fig:eta_2d_heatmap}).
      
    In the case of intermediate overshoot (Set 2), we found 283 models that have $n_\mathrm{pg} = 5$ excitation.
    The mass range spans from $6$ to $10.5\,M_\odot$ and metallicities from $Z= 0.0015$ to $0.0200$. 
    The most frequent values are $10\,M_\odot$ and $Z=0.0015$.

    On the other hand, in the no-overshoot models, mode $n_\mathrm{pg}=5$ excitation is relatively rare, with only 51 models showing instability. 
    These span a narrow mass range of $9- 12.5\,M_\odot$, with metallicities between $Z= 0.0015$ and $0.0055$; above this threshold, only isolated cases appear. 
    The most frequent configuration is $11.5\,M_\odot$ with $Z=0.0035$, indicating that excitation in this set favors high-mass, low-metallicity models.

    Overall, the results suggest that mode $n_\mathrm{pg}=5$  excitation favors low-metallicity and intermediate convective overshoot, with dominant masses typically between $9$ and $11\,M_\odot$ (see Figures\,\ref{fig:pos_etas} and \ref{fig:eta_2d_heatmap}).
    The figures also illustrate that increasing the overshoot parameter (from Set 1 to 3) leads to higher growth rates for the $n_\mathrm{pg} = 5$ mode, indicating that this mode becomes more strongly excited in models with extended convective mixing envelopes.
    Furthermore, a shift toward lower-mass models with increasing convective overshoot is also evident: while Set 0 is dominated by $11.5\,M_\odot$, Sets 1–3 show peak excitation at $9-10\,M_\odot$. 
    This trend is because of the narrowing of the mass range in which blue loop crossings can occur as overshoot increases, restricting mode excitation in low- and high-mass models.

\vskip 1em

\noindent\underline{\textbf{Mode $n_\mathrm{pg} = 6$}}
\vskip 0.5em

    \noindent We found a total number of 6945 models that have excitation of  $n_\mathrm{pg} = 6$ across the model grids. This makes $n_\mathrm{pg}=6$ the third most frequently excited pulsation mode in our model grid.
    Set\,0 includes 1266 cases, spanning a mass range of $9-12.5\,M_\odot$ and metallicities across the full investigated range between $Z=0.0015-0.0200$.
    Note, however, that excitation is absent in the metallicity range between $Z=0.0075$ and $0.0105$, creating a noticeable gap within the continuous distribution.
    This absence is consistent with the findings discussed in Section\,\ref{sec:set0_blueloops}, where we showed that stars in this metallicity regime do not undergo blue loop evolution, thus failing to reach the structural conditions required for strange mode excitation.
    The most frequent combination in Set 0 models is $10.5\,M_\odot$ at $Z=0.0030$.
    
    \noindent Sets\,1 and\,3 show markedly different distributions for mode $n_\mathrm{pg} = 6$ excitation.  
    Set\,1 contains 2301 excited models, spanning a broad mass range of $4.50$–$11.50\,M_\odot$ and covering the full metallicity interval from $Z = 0.0015$ to $0.0200$.  
    Excitation peaks at $10.5\,M_\odot$ and $Z = 0.0015$.
    
    In contrast, Set\,3 includes only 1042 excited models, with excitation confined to $6.00$–$9.50\,M_\odot$ and metallicities between $Z = 0.0015$ and $0.0185$.  
    While the mode appears at the lowest tested metallicities ($Z = 0.0015$–$0.0020$), excitation is absent in the intermediate range between $Z \sim 0.0060$ and $0.0075$, creating a discontinuity similar to that seen for $n_\mathrm{pg} = 5$ (see left-hand panel of Figure\,\ref{fig:eta_2d_heatmap}).

    The highest number of excited models (2285) was found in Set 2.
    In this case, the mass range spans from $5$ to $10.5\,M_\odot$ and the metallicity varies from $Z=0.0015$ to $0.0200$. 
    The most common combinations are $10.0\,M_\odot$ and $Z=0.0015$.

    Overall, we found that mode $n_\mathrm{pg}=6$ excitation favors low-metallicity, intermediate-mass models, with dominant masses between $8.5$ and $10.5\,M_\odot$. 
    Additionally, we found that intermediate convective overshoot (Set 2) yields the highest number of excited cases, while Sets 1 and 3 show nearly identical distributions (see Figures\,\ref{fig:pos_etas} and \ref{fig:eta_2d_heatmap}).

\vskip 1em

\noindent\underline{\textbf{Mode $n_\mathrm{pg} = 7$}}
\vskip 0.5em

\noindent With an occurrence of 13\,767 cases across all sets, mode $n_\mathrm{pg} = 7$ is the most frequently excited overtone in our simulations. 
We found 2817 models in the case of Set 0, spanning a broad mass range of $5-12.5\,M_\odot$ and metallicities between $Z=0.0015$ and $0.0200$. 
However, consistent with what was described in Section\,\ref{sec:set0_blueloops}, the metallicity range is separated into two distinct islands as mode excitation is absent between $Z=0.0075$ and $0.0105$ (see Figure\,\ref{fig:eta_2d_heatmap}).
The most frequent configuration is $M=10.5\,M_\odot$ with  a metallicity of $Z=0.0035$.

Our results reveal a substantial number of excited models in Set\,1, with a total of 4660 cases showing instability in mode $n_\mathrm{pg} = 7$.  
We found that excitation spans a broad mass range of $4.5$–$11.5\,M_\odot$, with a mean value of $9.25\,M_\odot$, and occurs across the full metallicity interval from $Z = 0.0015$ to $0.0200$.  
The most frequently excited configuration corresponds to $9.5\,M_\odot$ at $Z = 0.0015$.

Set\,2 contributes 4193 excited models with mode $n_\mathrm{pg} = 7$.  
The mass range extends from $5$ to $10.5\,M_\odot$.  
Excitation spans the full metallicity interval between $Z = 0.0015$ and $0.0200$.  
The most frequently excited configuration corresponds to $9.5\,M_\odot$ at $Z = 0.0085$.

Set\,3 includes 2097 excited models for mode $n_\mathrm{pg} = 7$, distributed over a narrower mass range of $6$–$9.5\,M_\odot$ and metallicities from $Z = 0.0060$ to $0.0200$ (see Figures\,\ref{fig:eta_2d_heatmap} and \ref{fig:pos_etas}).  
Furthermore, mode excitation also reappears at the lowest metallicities ($Z = 0.0015$–$0.0020$), leading to a gap in the intermediate metallicity regime.  
Mode $n_\mathrm{pg} = 7$ excitation peaks at $8.5\,M_\odot$ and $Z = 0.0100$ models.

\begin{figure*}
    \centering
    \includegraphics[width=1\linewidth]{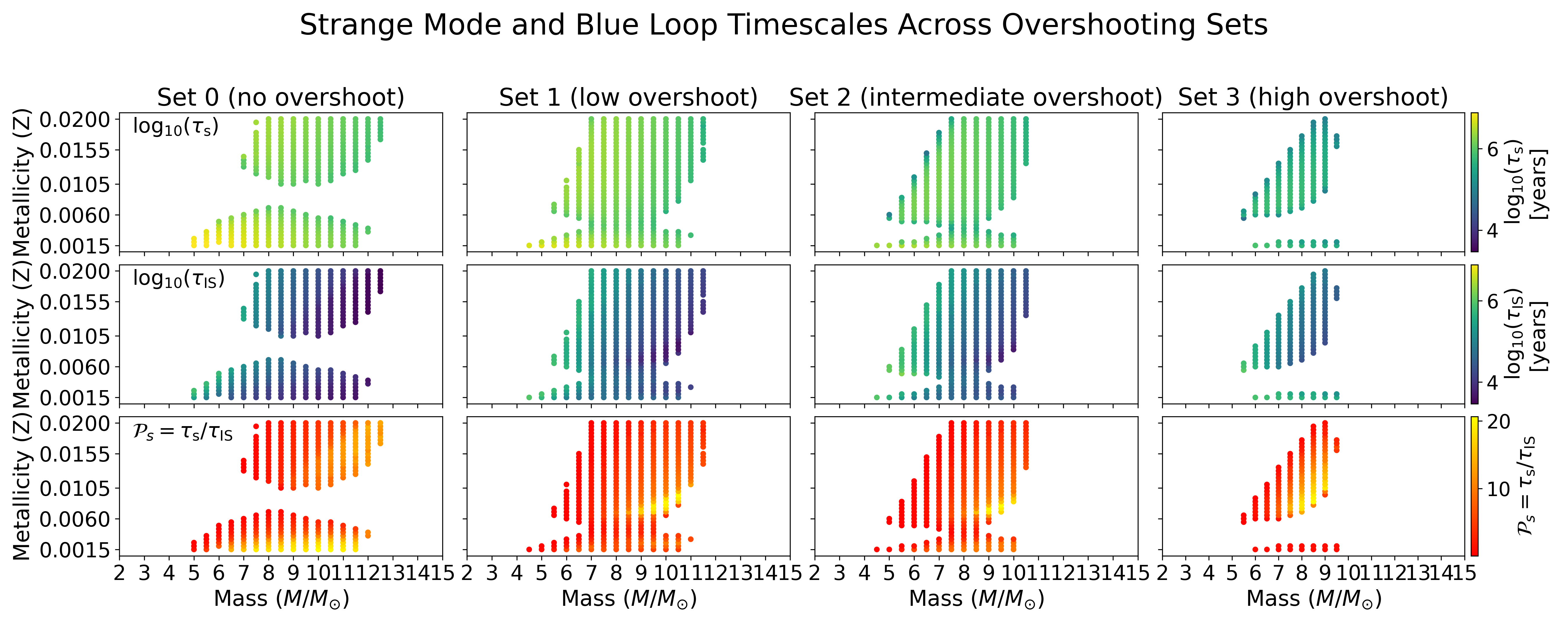}
    \caption{Scatter plots showing the timescales and their ratio for strange mode pulsations across all investigated stellar models. Each column corresponds to a different overshoot prescription (Sets\,0 - 3). Color indicates the timescales (top and middle rows, using a shared palette for $\log_{10}(\tau_{\mathrm{s}})$ and $\log_{10}(\tau_{\mathrm{IS}})$) and their ratio (bottom row, with a distinct palette for the dimensionless pulsation potential $\mathcal{P}_s = \tau_{\mathrm{s}} / \tau_{\mathrm{IS}}$). 
}
    \label{fig:timescales}
\end{figure*}
 
\vskip 1em

\noindent\underline{\textbf{Mode $n_\mathrm{pg} = 8$}}
\vskip 0.5em

Mode $n_\mathrm{pg} = 8$ appears in a total of 7058 models across the full grid, making it the second most frequently excited overtone mode in our simulations.  
Set\,0 includes 2548 excited models, spanning a wide mass range from $5$ to $12.5\,M_\odot$.  
Additionally, the metallicity range extends across the full investigated regime ($Z = 0.0015$–$0.0200$)
Although similar to $n_\mathrm{pg} = 6-7$ mode excitations, a gap can be seen between $Z = 0.0075$ and $0.0105$, as stars within this regime fail to develop blue loops on the HRD (see Section\,\ref{sec:set0_blueloops}).  
The most frequent configuration corresponds to $9\,M_\odot$ at $Z = 0.0200$.

In the case of Set\,1 models, we found 2479 excited models, with broad mass range of $4.5$–$11.5\,M_\odot$.
Furthermore, the metallicity range covers the range between $Z = 0.0015$ and $0.0200$.  
The most common configuration is $8\,M_\odot$ with $Z = 0.0200$ (see Figure\,\ref{fig:eta_2d_heatmap}).

Set\,2 contains 1742 excited models, distributed over a mass range of $4.5$–$10\,M_\odot$.  
Similar to lower radial modes, metallicities span over the full interval ($Z = 0.0015$–$0.0200$).
The most frequent configuration appears at $7.5\,M_\odot$ and $Z = 0.0180$ model.  

Considering Set\,3, we found the lowest number of excited models, with 289 cases. distributed between $5.50$ and $8.50\,M_\odot$ and metallicities from $Z = 0.0055$ to $0.0195$.  
The mean mass is $6.87\,M_\odot$, and excitation peaks at $7.00\,M_\odot$ and $Z = 0.0060$ (63 and 17 occurrences).  
Interestingly, mode excitation also reappears at the lowest tested metallicities ($Z = 0.0015$–$0.0020$), bypassing the intermediate regime.  
This fragmented behavior in metallicity suggests a narrow structural window for mode growth under low overshoot conditions.

\vskip 1em

\noindent\underline{\textbf{Mode $n_\mathrm{pg} = 9$}}
\vskip 0.5em

Mode $n_\mathrm{pg} = 9$ appears in only a few cases in our simulations, with a total of 10 excited models across the grid.  
Excitation is restricted only to Sets\,0 and\,1, and no higher-order modes beyond $n_\mathrm{pg} = 9$ were found to be excited at positive $\eta$ values in our models.

Set\,0 contains 9 excited models, spanning a narrow mass range of $9$–$10\,M_\odot$ and metallicities between $Z = 0.0175$ and $0.0200$.  
The most frequent configuration at $9.00\,M_\odot$ and $Z = 0.0195$.
Besides, in the case of Set\,1, we only found one, isolated model that has $n_\mathrm{pg} = 9$ excitation ($8.5\,M_\odot$ and $Z = 0.0195$).

\subsection{Lifetime of Strange Modes Across the Blue Loop}

    In order to quantify the detectability of strange modes, we calculated three key parameters for each relevant model, as defined in Section~\ref{sec:mathcalp}. 
    First, we determined the total timescale of the instability crossing phase ($\tau_\mathrm{IS}$), defined as the cumulative time spent within the instability strip. 
    This includes all intervals between entry at the red edge and exit at the blue edge, accounting for multiple re-entries from the blue side if present.

    Then we computed the duration of the strange mode pulsation phase ($\tau_\mathrm{s}$). Finally, we evaluated their ratio, the dimensionless pulsation potential $\mathcal{P}_\mathrm{s} = \tau_\mathrm{s}/\tau_\mathrm{IS}$, which captures the fractional coverage of the strange mode phase and the instability crossing phase within the blue loop.
    High values of $\mathcal{P}_\mathrm{s}$ indicate that strange mode instability lasts longer than the instability crossing phase throughout the blue loop phase, increasing the statistical likelihood of observing such pulsations. 
    In contrast, low values suggest that the instability is restricted mostly to a brief evolutionary window compared to the instability phase, reducing its detectability.
    Figure~\ref{fig:timescales} summarizes these three timescale diagnostics parameters. The top row shows $\log_{10}(\tau_\mathrm{s})$, the middle row $\log_{10}(\tau_\mathrm{IS})$, and the bottom row the dimensionless ratio $\mathcal{P}_\mathrm{s}$. 
    Columns from left to right correspond to the four overshoot prescriptions (Sets 0 - 3). 
    To ensure comparability between $\tau_\mathrm{IS}$ and  $\tau_\mathrm{s}$, a shared color palette is used.
    
    The results reveal that $\tau_\mathrm{s}$ spans approximately $\sim 10^{4.5}$ to $\sim 10^6 - 10^{7}$ years across all models. 
    The longest lifetimes are found in no-, or low-overshoot models (Sets 0 and 1), particularly at low mass and low metallicity. 
    On the other hand, $\tau_\mathrm{IS}$ covers a range of $\sim 10^4 - 10^6$ years, almost an order of magnitude lower than $\tau_\mathrm{s}$.
    Furthermore, in Figure\,\ref{fig:timescales}, we observe a strong trend: increasing both $Z$ and $M$ leads to a significant reduction in $\tau_\mathrm{IS}$.
    A similar pattern can be seen for the strange mode phase, also. 
    The longest $\tau_\mathrm{s}$ values also occur at low $M$ and $Z$, indicating that these conditions favor a long-lived strange mode pulsating phase.

    Considering the ratio $\mathcal{P}_\mathrm{s}$, we found that in Set 0 models, the ratio increases with $M$ in low-metallicity models (up to $\sim 15-20$).
    This means that, in these configurations, the star spends its lifetime more than an order of magnitude longer in strange mode phase than in the instability strip.
    However, increasing $Z$ reduces $\mathcal{P}$ (down to e.g., $\sim 2.5$ in $M = 7\,M_\odot$, $Z=0.0130$; or more). 

    Across all overshooting sets, the timescale ratio $\mathcal{P}_\mathrm{s} = \tau_\mathrm{s} / \tau_\mathrm{IS}$ remains mainly on the order unity or below, indicating that stars typically spend a comparable amount of time in the strange mode phase as in the instability strip itself. 
    Despite variations in the absolute durations of both phases, this balance is a robust feature across the grid. 
    However, a sharp increase in $\mathcal{P}_\mathrm{s}$ is consistently observed toward higher stellar masses at intermediate metallicities, independent of the overshoot prescription. 
    In all Sets, models with $M \gtrsim 8 –8.5\,M_\odot$ and $Z \approx 0.0060–0.0100$ exhibit an increase in the ratio, occasionally reaching values as high as $\mathcal{P}_\mathrm{s} \sim 10–20$. 
    These findings suggest that, while the strange mode phase is generally short-lived, it can dominate the blue loop phase in intermediate- to high-mass stars, depending on the convective structure of the star.
    
\begin{figure*}
    \centering
    \includegraphics[width=\textwidth]{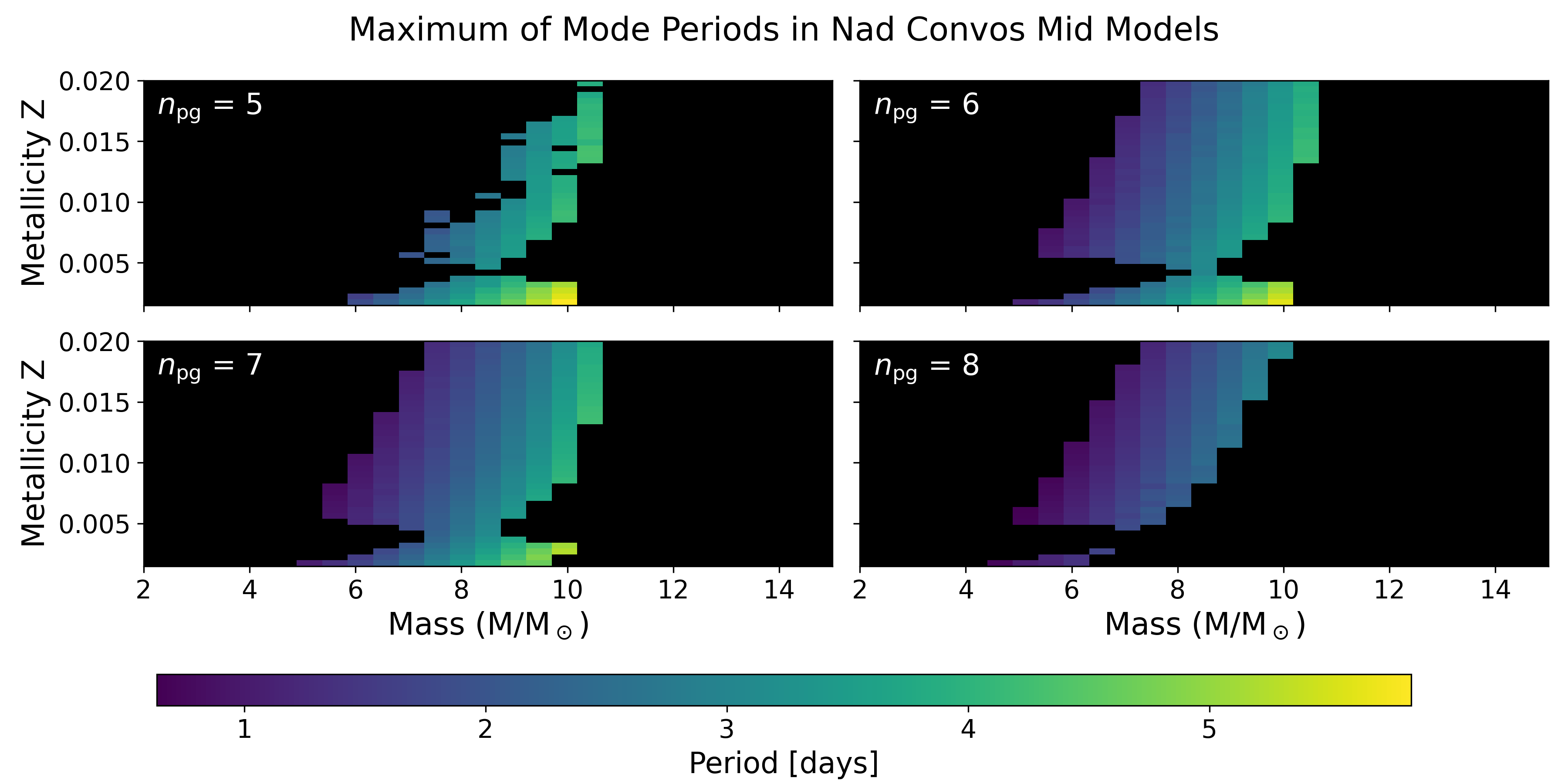}
    \caption{The maximum period of each radial order ($n_\mathrm{pg} = 5 -8$) on the $M-Z$ grid for the intermediate overshoot models (Set 2). For each subplot, the longest period associated with the given mode number is shown for a given $M-Z$ bin, selected across all \texttt{MESA RSP} models.}
    \label{fig:periods}
\end{figure*}

\begin{table*}
\centering
\caption{Period ranges are shown across overshoot sets on the $M$–$Z$ plane, organized by model set and radial order ($n_\mathrm{pg}$). For each period extrema, the corresponding mass and metallicity values are indicated. The absolute maximum period per overshoot set is highlighted with boldface.}

\begin{tabular}{cccc}
\hline
\hline
\textbf{Mode ($\mathbf{n_\mathrm{pg}}$)} & \textbf{Period range (days)} & \textbf{$\mathbf{P_\mathrm{min}}$ configuration} & \textbf{$\mathbf{P_\mathrm{max}}$ configuration} \\
\hline

    \hline
    \multicolumn{4}{c}{\textbf{Set 0 — No overshoot}} \\[1ex]
    \hline
5 & $2.80 - \mathbf{5.07}$ & 9.0 $M_\odot$, $Z=$0.0035 & 11.5 $M_\odot$, $Z=$0.0015 \\
6 & $0.77 - 4.86$ & 5.0 $M_\odot$, $Z=$0.0020 & 12.0 $M_\odot$, $Z=$0.0035 \\
7 & $0.66 - 4.77$ & 5.0 $M_\odot$, $Z=$0.0025 & 11.5 $M_\odot$, $Z=$0.0015 \\
8 & $0.62 - 4.03$ & 5.0 $M_\odot$, $Z=$0.0025 & 12.5 $M_\odot$, $Z=$0.0195 \\
9 & $1.26 - 1.76$ & 9.0 $M_\odot$, $Z=$0.0185 & 10.0 $M_\odot$, $Z=$0.0195 \\
    \hline
    \multicolumn{4}{c}{\textbf{Set 1 — Low Overshoot}} \\[1ex]
    \hline
5 & $1.84 - \mathbf{5.51}$ & 6.5 $M_\odot$, $Z=$0.0020 & 11.0 $M_\odot$, $Z=$0.0030 \\
6 & $0.74 - 5.22$ & 4.5 $M_\odot$, $Z=$0.0015 & 11.0 $M_\odot$, $Z=$0.0030 \\
7 & $0.68 - 5.22$ & 4.5 $M_\odot$, $Z=$0.0015 & 11.0 $M_\odot$, $Z=$0.0030 \\
8 & $0.64 - 3.61$ & 5.5 $M_\odot$, $Z=$0.0080 & 11.5 $M_\odot$, $Z=$0.0180 \\
9 & 1.35 & $-$ & 8.5 $M_\odot$, $Z=$0.0195 \\
    \hline
    \multicolumn{4}{c}{\textbf{Set 2 — Intermediate Overshoot}} \\[1ex]
    \hline
5 & $1.62 - \mathbf{5.84}$ & 6.0 $M_\odot$, $Z=$0.0020 & 10.0 $M_\odot$, $Z=$0.0015 \\
6 & $0.89 - 5.61$ & 5.5 $M_\odot$, $Z=$0.0075 & 10.0 $M_\odot$, $Z=$0.0015 \\
7 & $0.79 - 5.23$ & 5.5 $M_\odot$, $Z=$0.0080 & 10.0 $M_\odot$, $Z=$0.0025 \\
8 & $0.64 - 3.07$ & 5.0 $M_\odot$, $Z=$0.0060 & 10.0 $M_\odot$, $Z=$0.0190 \\
    \hline
    \multicolumn{4}{c}{\textbf{Set 3 — Strong Overshoot}} \\[1ex]
    \hline
5 & $1.78 - \mathbf{6.32}$ & 6.5 $M_\odot$, $Z=$0.0085 & 9.5 $M_\odot$, $Z=$0.0015 \\
6 & $1.67 - 6.08$ & 6.5 $M_\odot$, $Z=$0.0090 & 9.5 $M_\odot$, $Z=$0.0015 \\
7 & $1.26 - 3.96$ & 6.0 $M_\odot$, $Z=$0.0080 & 8.0 $M_\odot$, $Z=$0.0020 \\
8 & $1.09 - 2.56$ & 5.5 $M_\odot$, $Z=$0.0065 & 8.5 $M_\odot$, $Z=$0.0170 \\
\hline
\end{tabular}
\label{tab:period_extrema}
\end{table*}

\subsection{Period Ranges from Linear Stability Analyzis}

To complement the timescale-based analysis of strange mode lifetimes, we examined the period distribution of excited strange modes across the $M$–$Z$ grid. 
For each $M$–$Z$ combination, we ran multiple \texttt{MESA RSP} calculations, preselected in earlier steps. 
As the star evolves at fixed $M$ and $Z$, the excited strange mode may appear at different radial orders ($n_\mathrm{pg}$), with its period varying accordingly.

Figure\,\ref{fig:periods} displays, for each radial order ($n_\mathrm{pg} = 5$–$8$), the longest calculated period across all \texttt{MESA RSP} models within each $M$–$Z$ parameter space for Set 2. 
A shared colorbar is used to ensure direct comparability across panels. 
This maximum-based representation provides an upper bound for detectability, e.g., within a single TESS sector.

Our results reveal that across all overshoot prescriptions and radial mode orders, the period range spans approximately $0.6$ to $6.3$ days (see Table\,\ref{tab:period_extrema}). 
The absolute maximum values for each overshoot set are highlighted in boldface. 
Increasing the overshoot strength systematically broadens the period range at fixed $n_\mathrm{pg}$, with the longest periods consistently found in the strong overshoot models (e.g., $P = 6.32$ days in Set 3 for the $M = 9.5\,M_\odot$, $Z = 0.0015$ model).
Conversely, the shortest periods (down to $0.62$ days) appear consistently at low masses. 
Although Figure\,\ref{fig:periods} shows only the intermediate overshoot set, this mass–metallicity trend holds across all overshoot prescriptions.

However, it should be emphasized that all period values reflect linear stability predictions only, and do not imply actual mode excitation or observational detectability. 
Thus, the period calculations serve as a preliminary diagnostic, which must be further examined through nonlinear analyzis.

\section{Discussion and Conclusions}

   We now summarize the main results of this paper and outline key limitations of the applied methods.
   We also discuss the relevance of the detectability of strange modes and the need for future validation.

\subsection{Grid Construction and Mode Selection}

    To investigate the occurrence and detectability of strange mode pulsations, we constructed a dense grid of stellar evolution models using the stable version 23.05.1 of the open-source \texttt{MESA} software package. 
    We adopted four overshoot prescriptions (Sets 0–3) to explore the impact of convective mixing on the excitation of strange modes; the configurations are summarized in Table~\ref{tab:overshoot}.

    These configurations allowed us to investigate how the extent of convective penetration influences the extension and duration of the blue loop phase across the $M$–$Z$ grid. The complete grid contains 4104 individual simulations. 
    
    We filtered this grid to retain only models exhibiting a blue loop phase, using a custom pipeline built with \texttt{mesalab}.
    After filtering, we performed non-adiabatic pulsation analysis on each selected model using both \texttt{GYRE~7.0} \citep{Townsend2013,Townsendetal2018} and the \texttt{MESA RSP} module, focusing on the lowest 15 radial modes ($l=m=0$, $n_\mathrm{pg}=0$–$14$). 
    The \texttt{GYRE} simulations did not yield any positive growth rates above $n_\mathrm{pg}=2$, whereas the \texttt{MESA RSP} runs did reveal unstable modes beyond $n_\mathrm{pg}=2$.

\subsection{Strange Mode Excitation Patterns}

    We found that strange modes were typically well-separated from the fundamental and low-order overtones ($f_0$, $f_1$, $f_2$) and are highly sensitive to both stellar structure and the adopted convective overshoot prescription. 
    To isolate models with strange mode pulsations, we introduced a threshold at $n_\mathrm{pg}^\mathrm{th}=3$ and filtered those \texttt{MESA RSP} simulations that fulfilled this criterion. 
    Out of the full set of $135\,499$ \texttt{MESA RSP} runs, the fraction of models satisfying this criterion was $\sim 5\%$, $6.8\%$ $12.5\%$, and $\sim 10\%$ in Sets $0 - 3$, respectively, peaking at Set 2.
    These fractions indicate that strange mode instability is relatively rare across the full parameter space, but not negligible. 

    In the investigated parameter space, the excited modes predominantly appeared in the range $n_\mathrm{pg} = 5 - 9$, with no unstable modes detected beyond $n_\mathrm{pg} = 9$ in any simulation. 
    The most frequently excited modes were $n_\mathrm{pg} = 7$ and $8$, followed by $n_\mathrm{pg} = 6$. Modes at $n_\mathrm{pg} = 5$ occurred only sporadically, while $n_\mathrm{pg} = 9$ was the rarest among the detected unstable modes.
    This contrasts with earlier findings of \citet{Buchleretal1997}, who reported that strange modes are excited with $n_\mathrm{pg} =10- 12$ in a classical Cepheid. 
    However, simulations by \citet{BuchlerandKollath2001} showed that strange mode excitation can already be present at $n_\mathrm{pg} = 8-9$, which is consistent with our results.
    However, to our knowledge, no previous study has reported unstable strange modes at lower radial orders ($n_\mathrm{pg} \simeq 5-7$), as found in our simulations. 
    This suggests that the floor of strange mode excitation may be equally sensitive to chemical and structural properties of the star.

\subsection{Timescale Diagnostics}

    The results revealed a clear correlation between overshoot strength and the extent of the strange mode excitation domain. 
    As overshoot increases from Set~0 to Set~3, the instability region in both the $n_\mathrm{pg}$–$M$ and $n_\mathrm{pg}$–$Z$ planes became narrower. 
    In particular, increasing the overshoot envelope, strange modes tend to be shifted toward lower and intermediate-mass stars, and their occurrence became restricted to specific metallicity ranges at moderate - high values ($Z\ge 0.0060-0.0070$). 
    This is in good agreement with what was earlier found by \cite{Ziolkowska2024}.

    To quantify detectability, we introduced three timescale diagnostics: the absolute duration of the 
    crossing of the classical IS ($\tau_\mathrm{IS}$) and the strange mode instability phase ($\tau_\mathrm{s}$), and their dimensionless ratio, the potential of strange mode pulsation, $\mathcal{P}_\mathrm{s}$. 
    We found that increasing stellar mass tends to raise $\mathcal{P}_\mathrm{s}$, suggesting a higher likelihood of detection of a star in a strange mode unstable state. 
    However, increasing mass further tends to reduce the duration of the strange mode instability phase.
    Namely, from $\tau_\mathrm{s}\simeq10^{6}-10^{6.5}$ years to $\sim 10^{4.5}$–$10^{5}$ years, which is relatively a short period of time in the context of stellar evolution. 
    
    A clear mass dependence is only observed in the no-overshoot models (Set~0).
    Namely, at low metallicities, $\mathcal{P}_\mathrm{s}$ increases with stellar mass, reaching values $\sim15-20$ in high mass models.
    In addition, all overshooting sets (Sets~1–3) exhibit a region of higher $\mathcal{P}_\mathrm{s}$ values at intermediate metallicities ($Z \sim 0.0060$–$0.0100$) and masses above $\sim 8$–$8.5\,M_\odot$.
    Here, the ratio can reach values as high as $10–20$. 
    This domain appears as a distinct island in the parameter space and is the most extended in metallicity for the high overshoot models (Set~3), where it spans a broader $Z$ range at fixed mass.

\subsection{The Position of Strange modes on the HRD and the \textit{Gaia} CMD}

    We identified the position of strange mode pulsators on the Hertzsprung-Russell, and the \textit{Gaia} color-magnidute diagrams.
    We found that strange mode pulsators form a distinct and isolated island, clearly separated from the loci of classical pulsators such as DSCT, DCEP, SPB, and BCEP variables. 
    Their position is offset towards bluer $(M_\mathrm{BP} - M_\mathrm{RP})$ colours compared to the Cepheid population, with no apparent overlap. 

\subsection{Observability with TESS and Gaia}

    Our results show that the periods of excited strange modes span a range from approximately $0.6$ to $6.3$ days, depending on radial order, stellar mass, metallicity, and overshoot prescription. 
    The longest periods occur in high-mass, low-metallicity models with strong overshoot, while the shortest are consistently found at low masses. 
    As TESS sectors typically last 27 days, these period values fall well within this observation window. 
    Even a single sector would cover 5–15 pulsation cycles.
    This would be sufficient to determine periods and light curve characteristics accurately: rotational modulation signals with similar periods and with amplitudes at the millimagnitude level have been detected before \citep[see, e.g.,][]{Canto-2020,Kochukov-2021}. 
    
    The main difficulty TESS observations pose is the low angular resolution and thus potential blending with other sources, especially close to the galactic plane. 
    However, that can be mitigated by examining the properties of \textit{Gaia} sources within the TESS pixel apertures. 
    Furthermore, once \textit{Gaia} DR4 becomes available, all \textit{Gaia} sources will have epoch photometry from the nominal mission duration published. 
    This will allow us to disentangle true candidates from blends between a high-luminosity target and and fainter contaminating source.

\section{Conclusions and Future Work }

    Strange mode pulsations seem to be a structurally sensitive phenomenon, confined to specific regions of the $M$--$Z$--overshoot space. 
    Their detectability depends on a delicate balance between excitation conditions, evolutionary timescales, and convective mixing. 
    These findings provide a physically motivated framework for identifying strange mode candidates and guiding future observational efforts.

    Ultimately, confirmation of strange mode Cepheids is in the purview of observation, not theory. 
    However, more sophisticated modeling will help to constrain observability. 
    Nonlinear models will tell us the most likely \teff{}--$L$ parameter space where these instabilities can grow into pulsations within the linear instability regions, as well as predict expected amplitudes along with the periods. 
    From there, a targeted search will be possible, using data from the \textit{Gaia} mission to identify candidates, and existing and future space photometric missions to confirm their variability. 
    Assuming that these stars exist and can be found, they may provide us with new reference points along the blue loops to accurately map them beyond the IS.
    As such, they would represent an important test of the evolutionary models and internal physics that determine the precise extent of the loops. 

\subsection{Key Caveats}

    While the instability strip boundaries used in this study follow canonical values adopted in \texttt{MESA} tutorials and Cepheid literature, it is important to note that these limits can vary with pulsation mode and metallicity \citep[see, e.g.,][and the references therein]{Fernie1990, Fiorentinoetal2002,Deka-2024, Espinozaetal2024}. 
    The adopted parameter space provides a general estimate, sufficient for identifying high-overtone strange modes, which consistently appear beyond the blue edge of the strip. 
    This choice may affect the inferred duration of the blue loop phase, but does not alter the presence of strange mode instability itself.
    
    Furthermore, we note that all linear \texttt{MESA RSP} simulations were run using the default convective configuration, which differs from the value adopted in the main evolutionary models \citep[see, e.g., ][]{Dasetal2021,Dasetal2024,Dasetal2025a,Dasetal2025b}. 
    While $\alpha_\mathrm{MLT}$ affects the evolutionary track and structural properties in static MLT evolutionary models \citep{Ziolkowska2024}, and RSP employs a simplified time‑dependent convection (TDC) scheme, neither approach provides a realistic treatment of convection. 
    This discrepancy does not alter our conclusions, since mode selection and pulsation outcome are governed primarily by the nonlinear evolution.
    
    We also emphasize that the present analysis is based on linear stability results obtained with \texttt{MESA RSP}. 
    While this approach provides a reliable basis for identifying the onset of strange mode instability, it is not sufficient to investigate the nonlinear evolution of pulsation amplitudes, mode selection, or potential mode interactions. 
    In particular, determining which modes will ultimately develop into actual pulsation and what photometric and RV amplitudes they may reach, as well as identifying whether they transform into higher-, or lower-order modes, is essential and requires a dedicated nonlinear stability analysis. 
    This aspect is crucial for connecting theoretical predictions to observations.

\subsection{Future Directions}

Future studies should expand the parameter space explored here. 
In particular, opacity tables have a large impact on ionization zones and may strongly influence strange mode excitation, but in the present work we only adopted a single table. 
Exploring opacity variations alongside metallicity and overshoot will provide a more complete picture. 

Detectability studies are also needed to connect theoretical predictions to observational strategies. 
Nonlinear models will help identify the most likely parameter space where instabilities grow into pulsations, guiding searches in \textit{Gaia}, TESS, and future photometric missions, such as the Nancy Grace Roman Space Telescope. In case of Roman, the Galactic Bulge Time Domain Survey and Galactic Plane Survey programs will have the potential to identify potential targets too faint for TESS \citep{Roman-TAC-2025,Roman-GPS-2025}. 

A significant complication in searching for real-world examples of strange mode Cepheids is that we expect these stars to be part of the thin disk population, just as classical Cepheids are. 
Interstellar extinction is highly variable within the disk, therefore we expect potential targets in the \textit{Gaia} database to spread out along the reddening vector considerably. This can be mitigated by using, for example, three-dimensional dust maps \citep[e.g.,][]{Green-2019}. However, even the reddening law changes within the disk, which also needs to be taken into account \citep{Zhang-Green-2025}.

Finally, visualization and interpretability tools are becoming increasingly important. 
Recent community efforts, such as \texttt{custom\_colors}\footnote{\url{https://zenodo.org/records/16050218}} \citep{custom_colors}, demonstrate the calculation of synthetic photometry during stellar evolution using the \texttt{MESA} colors module. 
This test evolves a stellar model while computing bolometric magnitudes and synthetic magnitudes in multiple photometric filters.  
Such functionality is particularly valuable as it enables direct comparison between theoretical models and observed photometry, and facilitates the calculation of light curves and other observable quantities. 
It would therefore be highly beneficial to implement this capability directly into \texttt{mesalab}, ensuring consistent presentation of stellar evolution and pulsation results across future studies.

\begin{acknowledgments}
This research was supported by the `SeismoLab' KKP-137523 \'Elvonal grant of the Hungarian Research, Development and Innovation Office (NKFIH), and by the LP2025-14/2025 Lendület grant of the Hungarian Academy of Sciences. 
This research was supported by the International Space Science Institute (ISSI) in Bern/Beijing through ISSI/ISSI-BJ International Team project ID \#24-603 - “EXPANDING Universe” (EXploiting Precision AstroNomical Distance INdicators in the \textit{Gaia} Universe). 
M.J. gratefully acknowledges funding of MATISSE: \textit{Measuring Ages Through Isochrones, Seismology, and Stellar Evolution}, awarded through the European 
Commission's Widening Fellowship.
This research made use of NASA’s Astrophysics Data System Bibliographic Services.

The simulations were performed on the high-performance computing machine at HUN-REN CSFK CSI acquired through the EC Horizon2020 project OPTICON (Grant Agreement No. 730890).

This paper was written with the assistance of a large language model (e.g., Gemini by Google, Microsoft Copilot, and ChatGPT by OpenAI) for linguistic  refinement. 
All content and any errors remain the sole responsibility of the first author.
\end{acknowledgments}

\begin{contribution}
D. T-N. performed the simulations, conducted the analysis, and wrote the manuscript.  
In addition, she performed the filtering and preparation of \textit{Gaia} CMD data used for comparative visualization.
L. M. provided conceptual guidance, scientific background, and overall supervision throughout the project. 
His input shaped the theoretical framework and informed the initial research direction. 
M. J. contributed to early-stage code development and offered technical input during the initial experimental design.
She also contributed to the final editing of the manuscript and offered technical guidance.

\end{contribution}

\software{Astropy \citep{astropy:2013, astropy:2018, astropy:2022}, F90nml \citep{f90nml}, GYRE \citep{Townsend2013}, H5py \citep{h5py}, Isochrones \citep{isochrones}, Matplotlib \citep{matplotlib}, MESA \citep{Paxton2011,Paxton2013,Paxton2015,Paxton2018,Paxton2019,Jermyn2023}, MESA RSP \citep{Paxton2019}, Mesalab \citep{TarczayNehezd2025},
Numpy \citep{numpy},
Pandas \citep{pandas}, Python \citep{Python3}, PyYAML \citep{pyyaml}, Seaborn \citep{seaborn}, Tqdm \citep{tqdm}}



\bibliography{references}{}
\bibliographystyle{aasjournalv7}

\pagebreak

\appendix

 \section{The \textit{Gaia} color-magnitude diagram for the selected blue loop crossing models}
\label{sec:appendixHRD}

\renewcommand{\thefigure}{A\arabic{figure}}
\setcounter{figure}{0}

 Figure\,\ref{fig:combined_cmd_filt} shows the models on the \textit{Gaia} color-magnitude diagram plane that underwent filtering via \texttt{mesalab} and were found to exhibit a blue loop phase, as required by our selection criteria.

 \begin{figure*}[h!]
    \centering
    \includegraphics[width=0.95\textwidth]{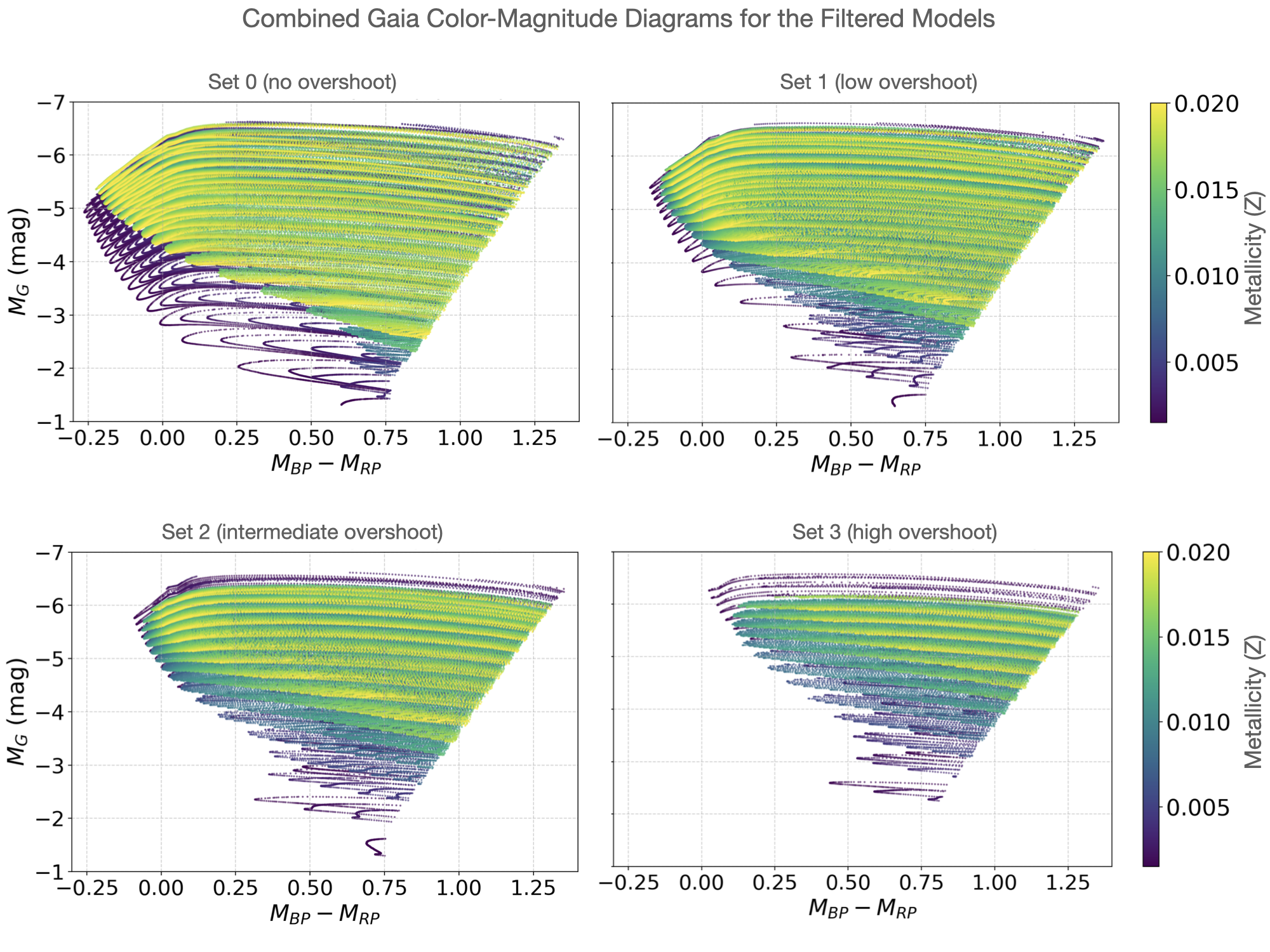}
    \caption{Similar to Figure\,\ref{fig:combined_hrd_filt}, but projected on the \textit{Gaia} $M_\mathrm{G}$ vs ($M_\mathrm{BP} - M_\mathrm{RP}$) plane for all overshoot prescriptions (Sets\,0 - 3). As seen in Figure\,\ref{fig:combined_hrd_filt}, color-coding denotes the value of $Z$ metallicity.}
    \label{fig:combined_cmd_filt}
\end{figure*}

\section{Structural Signatures of Strange Modes: Two Representative Models}
\label{app:strange_mode_struct}

\renewcommand{\thefigure}{B\arabic{figure}}
\setcounter{figure}{0}
    Figures\,\ref{fig:eigen_work_app1} and \ref{fig:eigen_work_app2} show the radial displacement and work integral profiles for two selected models that meet the $n_\mathrm{pg} > n_\mathrm{pg}^{th}$ criterion described in Section\,\ref{sec:strange_mode_selection}. 
    Both models were run with a moderate convective overshoot envelope prescription (Set 2).
    In Figure\,\ref{fig:eigen_work_app1}, the star has a mass of $5\,M_\odot$ and a metallicity of $Z=0.0015$ (with a model number of 1848), while in Figure\,\ref{fig:eigen_work_app2} these parameters are set to $M=8.5\,M_\odot$ and $Z=0.0195$ (with model number of 2183).

    In both cases, the strange mode is clearly distinguishable from the other modes (marked with a red line and $\mathcal{S}$ in the top left corner of the subplot).
    In the first model (Figure\,\ref{fig:eigen_work_app1}), it corresponds to the fifth overtone, while in the second (Figure\,\ref{fig:eigen_work_app2}), it appears as the ninth order. 
    The radial displacement profile shows a sharp localization in the outer layers in both cases (see upper panels), suggesting a radial pulsation in the uppermost layers (separated from the interior of the star).
    Considering the work integral of these stars, these figures show a concentrated driving region with the only positive net work integral in the case of strange modes. 
    These structural features confirm that the threshold-based selection effectively isolates strange mode pulsations of the stars.
    
\begin{figure*}
    \centering
    \includegraphics[width=\textwidth]{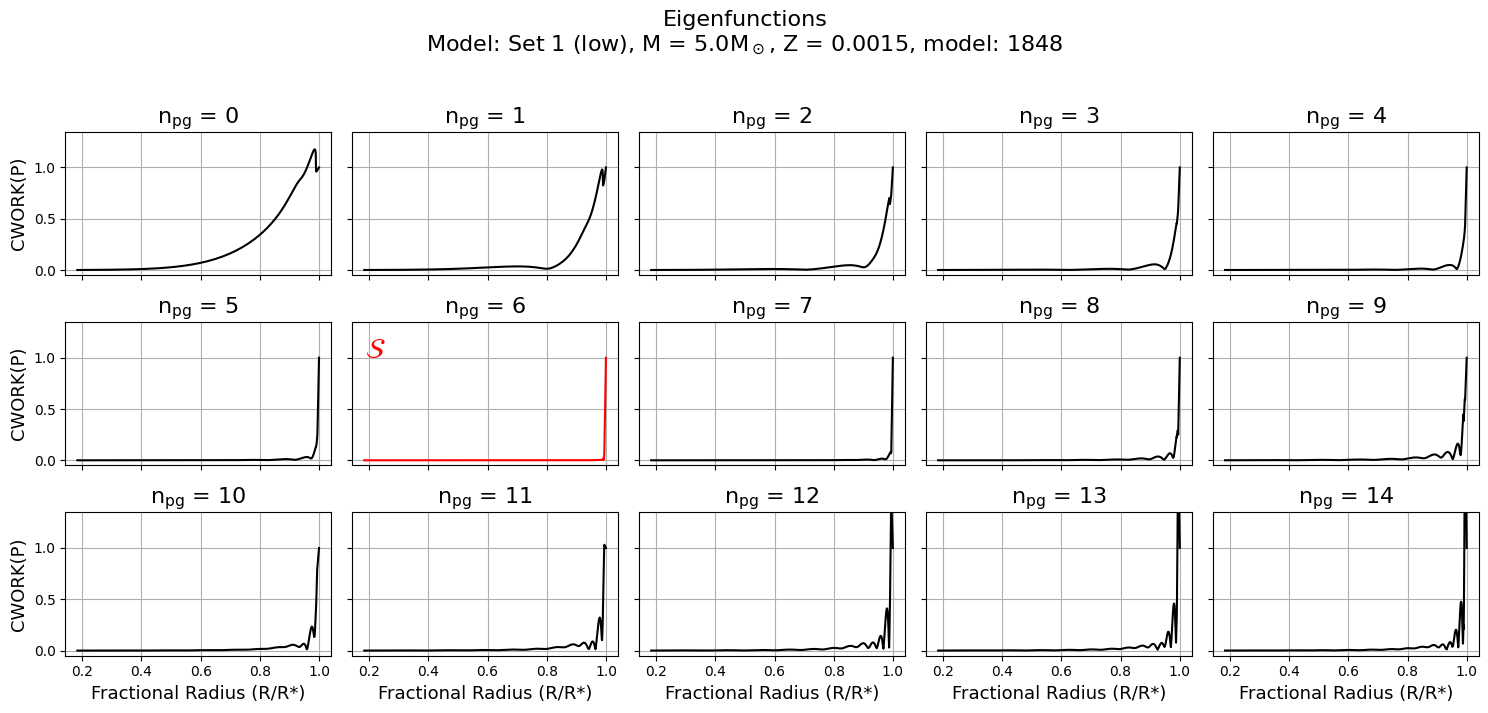}
    \includegraphics[width=\textwidth]{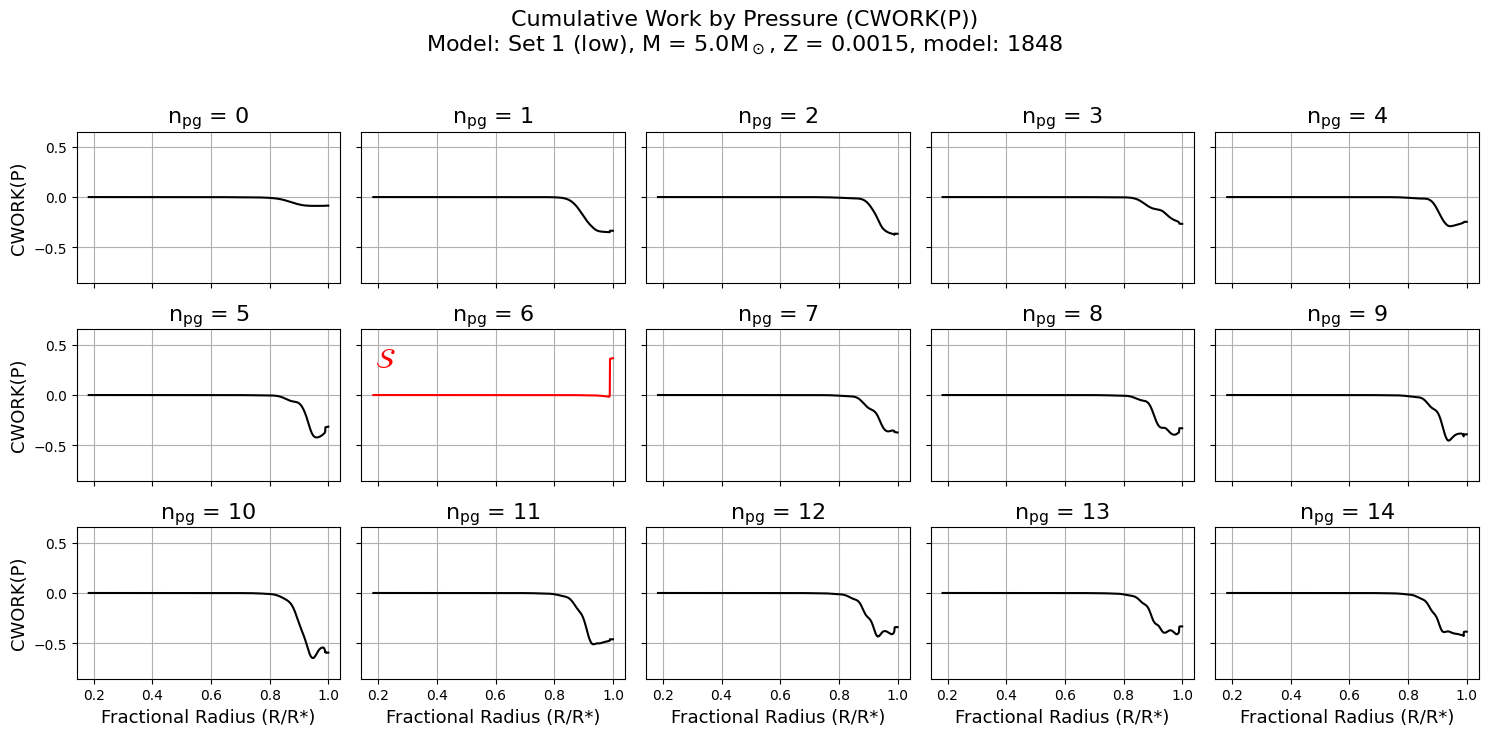}
    \caption{The radial displacement (upper panel) and work integral (lower panel) profile of a $5\,M_\odot$ star with a $Z=0.0015$ metallicity and moderate overshoot envelopes (Set 2). The star is found to pulsate only in the sixth overtone ($n_\mathrm{pg}=6$, marked with a red line and $\mathcal{S}$ in the upper left corner). The radial displacement eigenfunction shows a sharp discontinuity in the vicinity of the star's surface. Furthermore, this mode is the only one that is characterized by a net positive work integral ($\int \delta P \mathrm{d}\delta V>0$), which quantitatively confirms its linear instability and the strange mode nature of the pulsation.}
    \label{fig:eigen_work_app1}
\end{figure*}

\begin{figure*}
    \centering
    \includegraphics[width=\textwidth]{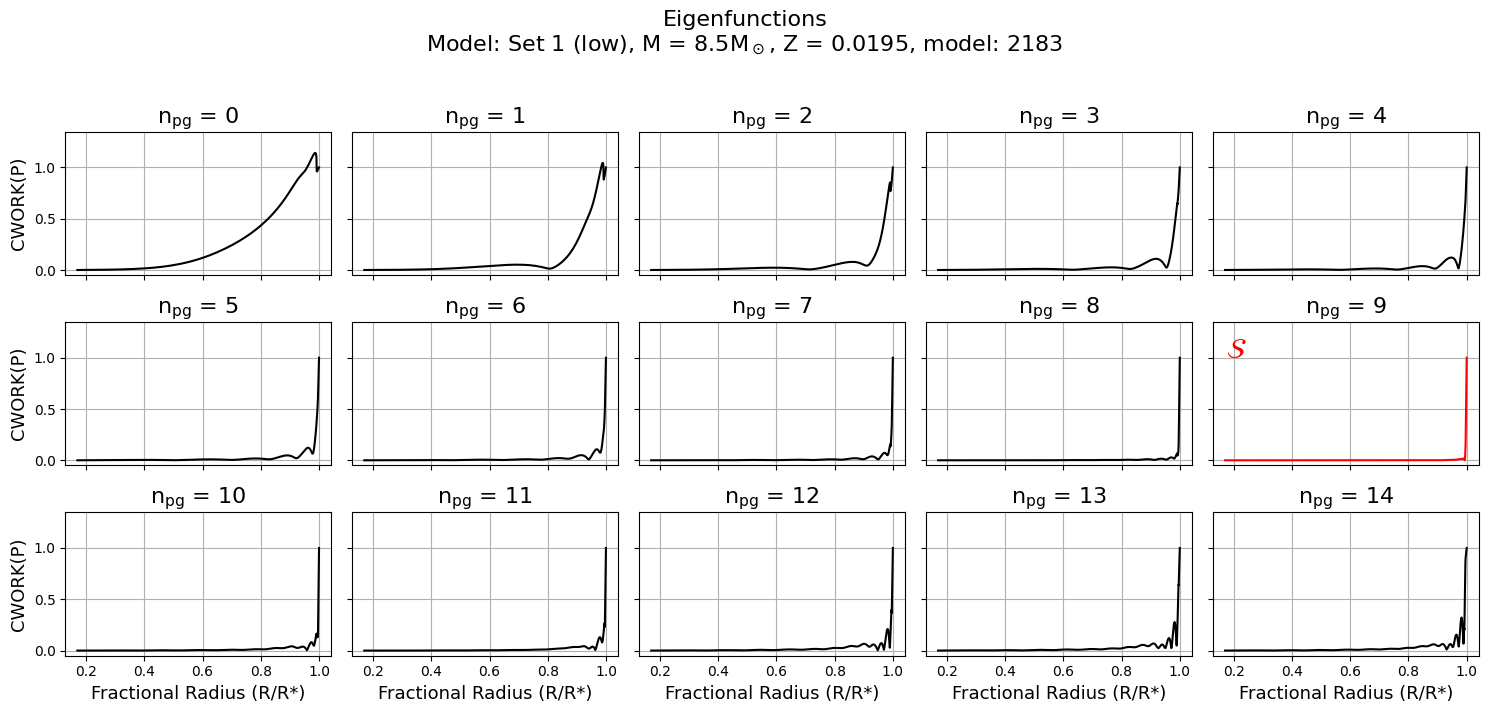}
    \includegraphics[width=\textwidth]{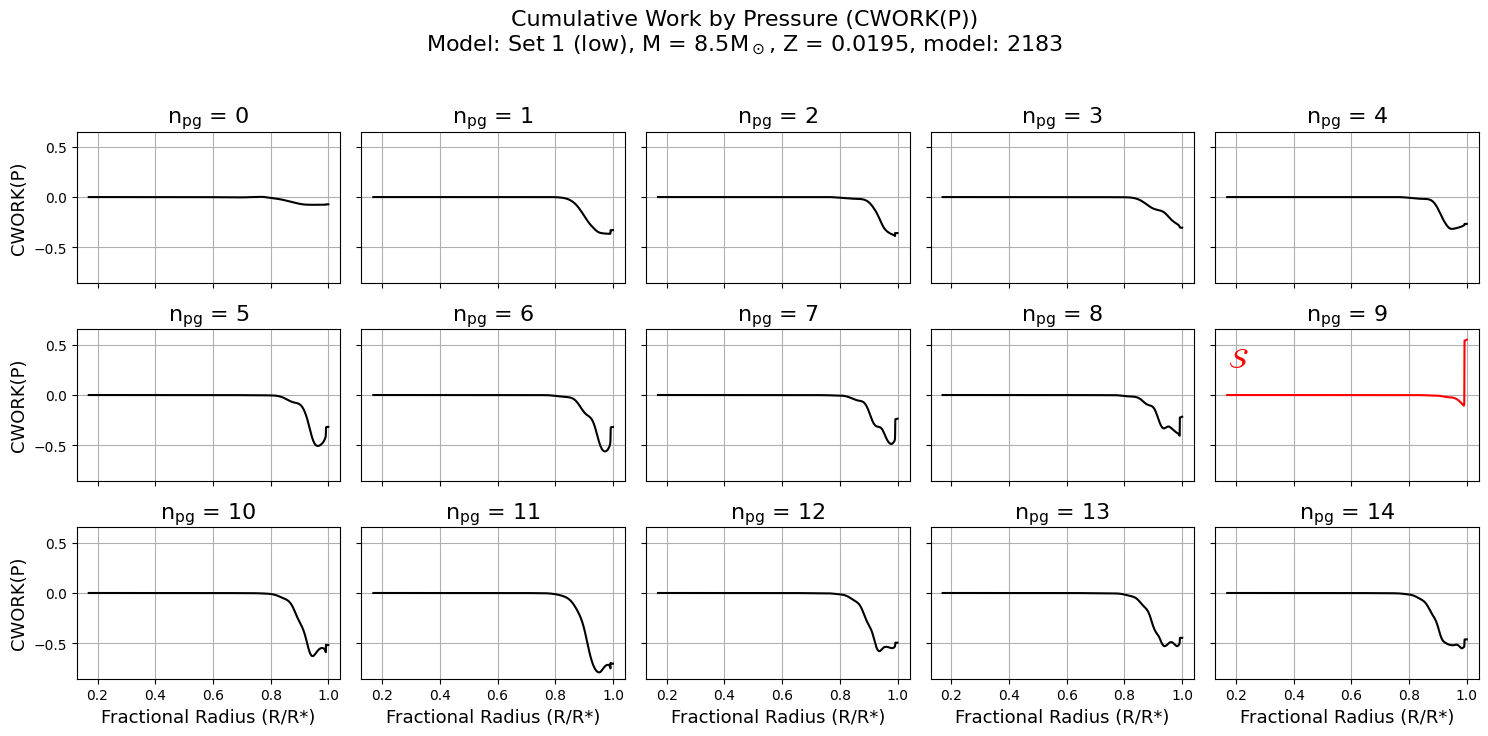}
    \caption{The radial displacement (upper panel) and work integral (lower panel) profile of an $8.5\,M_\odot$ star with a $Z=0.0195$ metallicity and moderate overshoot envelopes (Set 2). The star is found to pulsate only in the sixth overtone ($n_\mathrm{pg}=9$, marked with a red line and $\mathcal{S}$ in the upper left corner). The radial displacement eigenfunction shows a sharp discontinuity in the vicinity of the star's surface. Furthermore, this mode is the only one that is characterized by a net positive work integral ($\int \delta P \mathrm{d}\delta V>0$), which quantitatively confirms its linear instability and the strange mode nature of the pulsation.}
    \label{fig:eigen_work_app2}
\end{figure*}



\end{document}